\documentclass[10pt,twocolumn,floatfix]{revtex4-1}

\usepackage{amsmath}
\usepackage{amssymb}
\usepackage{graphicx}
\usepackage{color} 
\usepackage{multirow}
\usepackage[caption=false]{subfig}
\usepackage[normalem]{ulem}

\newcommand{\Pk}{$\text{P}[\mathbf{k}]$}
\newcommand{\Ppw}{$\text{P}[p(\mathbf{w})]$}
\newcommand{\Piso}{$\text{P}[\text{\textbf{iso}}(\mathbf{\Gamma})]$}

\graphicspath{{figures/}}

\begin{document}

%100 characters max
\title{Building connections: How scientists meet each other during a conference}
\date{\today}
\author{Mathieu G\'enois}
\email{mathieu.genois@gesis.org}
\affiliation{CNRS, CPT, Aix Marseille Univ, Universit\'e de Toulon, Marseille, France}
\affiliation{GESIS, Leibniz Institute for the Social Sciences, K\"oln/Mannheim, Germany}
\author{Maria Zens}
\affiliation{GESIS, Leibniz Institute for the Social Sciences, K\"oln/Mannheim, Germany}
%\author{Haiko Lietz}
%\affiliation{GESIS, Leibniz Institute for the Social Sciences, K\"oln, Germany}
\author{Clemens Lechner}
\affiliation{GESIS, Leibniz Institute for the Social Sciences, K\"oln/Mannheim, Germany}
\author{Beatrice Rammstedt}
\affiliation{GESIS, Leibniz Institute for the Social Sciences, K\"oln/Mannheim, Germany}
\author{Markus Strohmaier}
\affiliation{GESIS, Leibniz Institute for the Social Sciences, K\"oln/Mannheim, Germany}
\affiliation{RWTH Aachen University}

\begin{abstract}
%150 words max
We present the results of two studies on how individuals interact with each other during a international, interdisciplinary scientific conference. We first show that contact activity is highly variable across the two conferences and between  different socio-demographic groups. However, we found one consistent phenomenon: Professors connect and interact significantly less than the other participants. We interpret this effect as non-tenured researchers using conferences to accumulate social capital, while established researchers already have such capital. We then show that groups mix well during conferences, but note that a language-based homophily is always present. Finally, we show that the dynamics of the contacts across days is also similar between conferences. First day connections are established, then filtering occurs during the following days. The connection turnover between consecutive days proves to be large ($\sim$50\,\%), and related to the intensity of interactions.
\end{abstract}

\maketitle

%6000-7000 words, 8 Fig/Table max, ~50 ref
\section{Introduction}

%\hl{My five cents: In general, I like the main story, the idea that communication serves a function -- the networking function (Maybe use "networking" in the title?). The (naive) alternative would be a simple mechanistic model like the combination of the Matthew Effect and multi-homophily (Fariba's paper), according to which professors (high prestige) attract connections. For this you don't find any evidence. Maybe you could prominently use Merton's paper on the Matthew Effect to address this paradox (if it is one) and build the story around that. }
%\mz{I would like to point at the fact that Merton is generally seen as one of the (god)fathers of functionalism and, of course, I am more than willing to cite Merton (but maybe in a somewhat different way - imho the Matthew effect means that high reputation leads to more ("undeserved") reputation and concentration of resources; I do not see why high reputation should necessarily lead to more communication in that perspective and "attraction" is not "action". In fact I think one of our findings is that communication is a function of "status needed" rather than "status gained" -- I also think we overemphasise the status-argument and should rephrase these parts a bit. I do not quite understand what "multi-homophily" means?}

We investigate the temporal, disciplinary, and communicative structures and dynamics of interactions among scholars during conferences. We take advantage of the developments of methods that measure human behaviour \emph{in situ} to focus on empirical evidence of social interactions between individuals. These methods have been used to study many different social contexts: schools \cite{Salathe:PNAS2010,Sthele:PLOS2011,Guclu:PLOS2016}, hospitals \cite{Isella:PLOS2011,Hornbeck:JID2012,Vanhems:PLOS2013,Hertzberg:SN2017,Duval:SciRep2018,champredon:AIMS2018}, workplaces \cite{Genois:NS2015,Montanari:PerCom2017}, households \cite{Kiti:EPJDS2016,Ozella:PLOS2018}, conferences \cite{Hui:2005PSN,Barrat:ISWC2010,Atzmueller:MMUSM2012,Isella:JTB2016,Macek:2012ACM}, malls \cite{Ren:EPJDS2018} or even larger setups \cite{Stopczynski:PLOS2014}. Different technologies exist. Here we use the RFID-based system developed by the SocioPatterns collaboration \cite{Cattuto:PLOS2010}, which allows for the detection and collection of face-to-face physical proximity between individuals ($\sim$1.5\,m) with a high temporal resolution (20\,s interval). In addition to contact data, we also gather socio-demographic information about the participants through a survey, including their seniority and disciplinary background, country of residence, mother tongue, age, gender, and their roles at the conference (e.g. speaker, poster presenter, non-presenting participant, staff). The data were collected during two conferences organised in Cologne: the GESIS Computational Social Science Winter Symposium 2016 (WS16) and the International Conference on Computational Social Science 2017 (ICCSS17).

Our main interest is to mine patterns of contact behaviour (intensity and frequency) and relate them to the temporal process of the conference, \emph{i.e.} the different formats of networking slots (e.g., poster session, lunch break) on the one hand, and to the socio-demographic features of the attendees on the other. We uncover three main results on how researchers connected during these events. First, we find unambiguously that established and presumably tenured researchers --- professors --- connect and interact significantly less than average, despite being present as long as other groups of participants. At the same time, early to mid-career (and presumably non-tenured) researcher categories show either average or above than average connectivity and interactivity. We speculate on a possible mechanism to explain this behaviour, linked the use of conferences for capital accumulation . Second, we show that overall, participants mix rather well according to all socio-demographic dimensions. However, a small but detectable signal of language and country-based homophily appears. Finally, we investigate the dynamics of connections from one day to the next, and show the existence of a large turnover, along with a clear process of exploration and filtering of social interactions.

These results relate to research aspects from various fields, notably because we combine different classes of data (sensor data and survey data), computational methods and concepts taken from the social sciences, and, not the least, because of the object of study itself. The two conferences are indeed rooted in Computational Social Science, which is an emerging field of research. Our results also participate in understanding how a new paradigm appears in science.

In a sociology of science perspective, our results describe aspects of research collaboration \cite{katz1997research} \cite{toral2011exploratory} and the formation of a new research paradigm with regard to interdisciplinarity \cite{cummings2008collaborates}, new research ``cores'', disciplinary cohesion \cite{toral2011exploratory} \cite{moody2004structure} or dissolution \cite{abbott2010chaos}, emergence of new sub-disciplines and disciplinary identity \cite{lungeanu2015effects, darbellay2015rethinking}. The behavioural data we mine from the participants' interactions helps to find hypotheses on what is regarded ``functional'' \cite{labrie2015strategies} or ``pragmatic'' \cite{melin2000} in this specific interaction.
%\off{We identify collaboration trends (e.g. for or against internationalisation \cite{leydesdorff} or interdisciplinarity \cite{vanRijnsoever:collaboration}), clusters of interaction, brokerage positions \cite{collet2014} or the communication behaviour of area authorities. Further to that and on a rather abstract level our work contributes to the long established field of ``small group research'' \cite{moreland1982socialization} \cite{moreland1985new}, and, more specifically, to the relation between ``small groups'' and the more recent perspective on small groups and networks \cite{katz2004network}. Within the setting of an academic conference we observe the formation of (very) short-lived ``functional groups'' or ``ad hoc-groups''. The behavioural data we mine from the participants' interactions helps to find hypotheses on what is regarded ``functional'' \cite{labrie2015strategies} or ``pragmatic'' \cite{melin2000} in this specific  interaction: who talks with whom in what intensity and frequency. The development of communication over the two-day time-span of the conference is of particular interest, because we see which interactions persist and are intensified or, on the contrary, which are abandoned or avoided in the first place.}
%-------------------

In order to interpret our results, we draw upon the concept of social capital as developed in the works of French sociologist Pierre Bourdieu, which started with the Theory of Practise in 1972 \cite{bourdieu1972esquisse}. Bourdieu sees the (various) forms of capital as a societal \textit{movens}, as the ``\'energie de la physique sociale'',  that impacts the formation of social structure, individual behaviour, and group identities \cite{bourdieu1980capital}. Notions of social capital have been widely adopted since, not only in a Bourdieuan understanding; influential concepts have also been adapted to rational choice theory \cite{coleman1988} or implemented to capture the function of trust in networks \cite{putnam1995}. We relate, though, on the multidimensional concept of --- social, cultural, economic, and symbolic --- capital embedded in social theory and the analysis of the interactive struggles that take place in various social arenas or fields, the academic being one of them \cite{wacquant1990sociology} \cite{angervall2018academic} \cite{jungbauer2013determinants}. With his concept Bourdieu incorporates a relational perspective on human behaviour and the resources an individual can make use of and also allows for different (micro-meso-macro) levels of analysis. The (possible) tension between the ``position'' of an actor (the ``capital'' of institution, seniority, status of discipline, role at the conference etc.) and his or her "positioning" (the communicative outreach or containment) at the conference can be analysed in accordance with this sociological framework. This adds to the investigation on the role collaboration plays for the acquisition and deployment of capital, which in our case can be specified as scientific or academic capital. With our empirical data we shed light on the actual behaviour of academic crowds and can test hypotheses on motives for academic networking \cite{bozeman2004capital} and career strategies \cite{labrie2015strategies}. 

Our research also relates to previous scientometric work on co-authorship or citation networks that portrays structures of academic collaboration or knowledge trails across disciplines, the impact of gender, status or nationality on publication performance, and the effects of interdisciplinarity \cite{Newman2001structure}\cite{lewis2012disciplinarydifferences}\cite{milojevic2012academicage}\cite{vanRijnsoever:collaboration}\cite{leydesdorff}. However, we go beyond this approach in two aspects: co-authorship graphs focus on collaboration as evidenced by a publication, i.e. the visible result of a successful interaction. We instead look at the ``making-of'' and pre-history of what might become a collaboration of that kind or not. Moreover, \cite{katz1997research} have shown that many relevant sub-scenarios of research collaboration are not detectable via co-authorship analyses. The speedy and ad-hoc encounters and the selection processes going on during a conference highlight the choices taken and on different strategies of networking. We concentrate on the process of collaboration-making and are interested in choices and the explanation of choices or non-choices (strategies and the multitude of options rather than the result).

\section{Results}

\subsection{Group heterogeneities in interaction behaviour}\label{sec:group_hetero}

\begin{figure*}
  \centering
  \subfloat[\textbf{WS16}\label{fig:groups_WS16}]{
    \includegraphics[width=0.95\columnwidth]{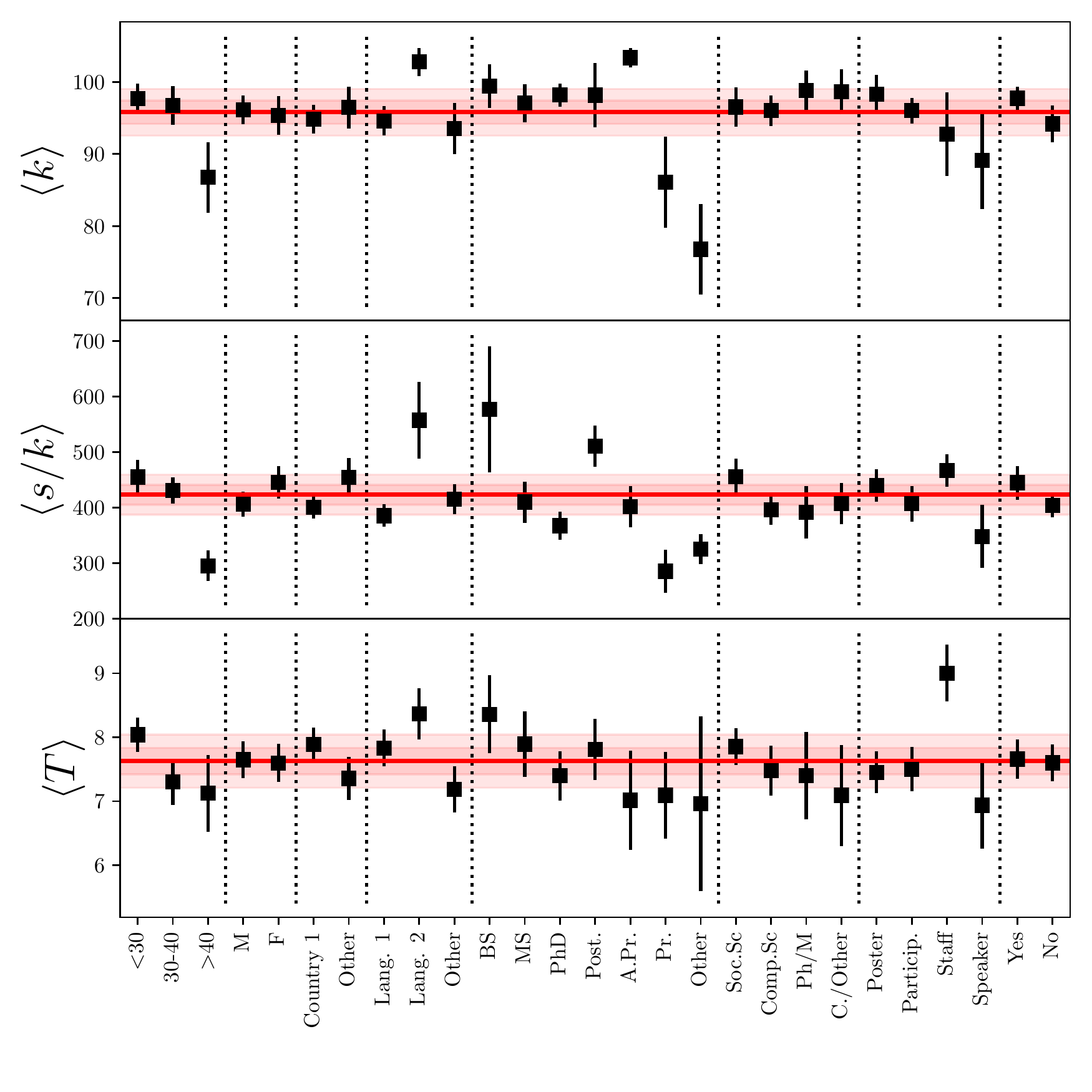}
  }
  \subfloat[\textbf{ICCSS17}\label{fig:groups_ICCSS17}]{
    \includegraphics[width=0.95\columnwidth]{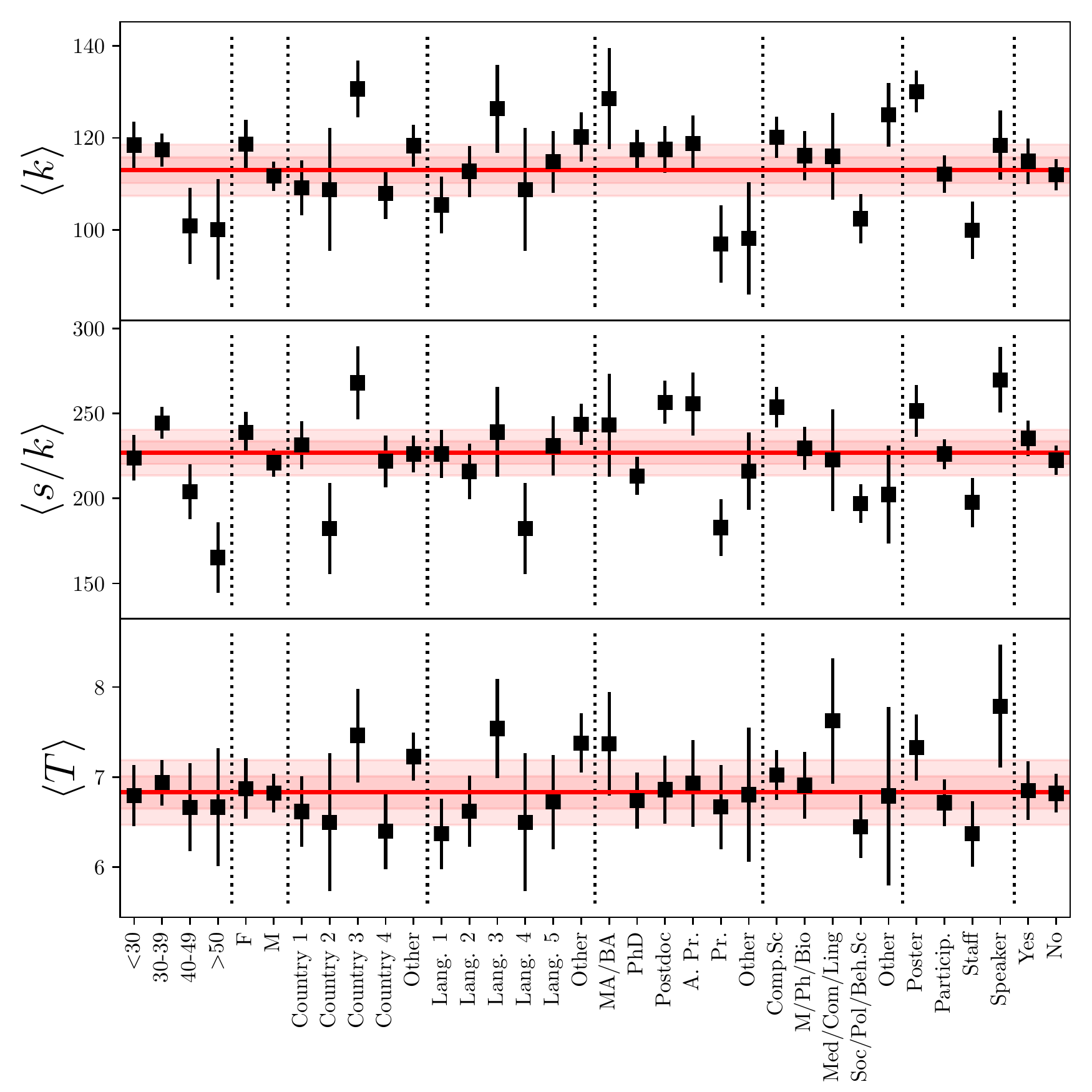}
  }
  \caption{\textbf{Differences in interaction behaviour between groups.} For each group we compute the average degree $\left<k\right>$, the average total contact duration per neighbour $\left<s/k\right>$ and the average total presence duration $\left<T\right>$ of the individuals, for the whole conference. The error bar shows the standard error on the measure. The solid coloured lines and coloured regions show the overall average value for all nodes, with one and two standard deviations.}
  \label{fig:groups}
\end{figure*}

The data collected using the SocioPatterns platform consists in network snapshots taken every 20 seconds, in which nodes are the participants and links are the contacts occurring during the snapshot. In other words, it is a temporal network with a resolution of 20\,s, in which links are present only when a contact occurs between two participants. As a remark, let it be clear that we call ``contact'' a face-to-face, physical proximity (less than 1.5\,m), as detected by the sensors from the SocioPatterns platform. In parallel, we collected the following information about the participants via a survey: Age group, gender, country of residence, mother tongue, academic seniority, disciplinary background, role in the conference, participation in a previous edition of the same conference (see SI for detailed information about the attributes).

We first focus on the aggregated network of interactions during the conferences, \emph{i.e.} we build the network where nodes are the participants and links exist between two participants if they have been in contact at least once during the whole event. Each link has a weight, defined as the total time each pair of participant has spent in contact. In this network, the degree of a node --- the number of participants with whom it has interacted --- measures the connectivity of the individual. A high degree means that the individual has connected with many participants during the conference. In order to measure the interactivity, \emph{i.e.} the average intensity of these connections, we compute for each individual the average interaction duration, \emph{i.e.} the average total time spent with each person he/she has been in contact during the whole conference.

% \begin{figure}
%   \centering
%   \subfloat[\textbf{WS16}\label{fig:k_groups_full_WS16}]{
%     \includegraphics[width=0.95\columnwidth]{k_groupes_full_WS16}
%   }
%   \\
%   \subfloat[\textbf{ICCSS17}\label{fig:k_groups_full_ICCSS17}]{
%     \includegraphics[width=0.95\columnwidth]{k_groupes_full_ICCSS17}
%   }
%   \caption{\textbf{Average degree in each group.} For each group we compute the average degree of the individuals, for the whole conference. The error bar shows the standard error on the measure of this average degree. The solid coloured line and coloured regions show the overall average value for all nodes, with one and two standard deviations.}
%   \label{fig:k_groups}
% \end{figure}

The contact networks are dense, in the sense that a large fraction of the possible links exist ($\rho = 0.793$ for WS16, $\rho = 0.507$ for ICCSS17), which leads to degree distributions that are skewed to the right side (see SI). This is the consequence of both conference venues being rather crowded during the events. Comparing the average degree within each group as defined by the socio-demographic attributes to the global average degree (Fig.~\ref{fig:groups}, top row) shows that significant differences in behaviour exist between these groups at both conferences. While for most the average degree is close to the global one, some groups exhibit either significantly lower or higher average degree. These abnormally low or high interaction groups are different across conferences, except for three negative outliers: Individuals over 40 years of age, professors, and ``other'' academic status groups. The latter is understandable, as this group is comprised of conference staff (and as such similar to the ``staff'' group with regards to  conference role) who are not part of the academic system and therefore constitute a separate, isolated group at the conference that is not expected to share interest with the general crowd, and thus interacts much less. However, the behaviour of older ($> 40$) participants and professors --- which are essentially the same group --- is surprising. A common model for interactions in a crowd in which a status hierarchy exists usually considers that individuals with higher status ``attract'' others with lower status. It appears, however, that in the contexts of the two scientific conferences, the opposite happens, or, at least, that attraction towards senior members does not translate into action. Note that the low-connectivity effect does not translate fully to the ``Speaker'' group. For WS16, speakers indeed show lower connectivity and interactivity, as expected since speakers are usually established researchers and hence high-status participants. However, we see a different picture at ICCSS17 where speakers interact more than average. This discrepancy can be explained: at ICCSS17, the "Speaker" category comprised not only keynote speakers but also a large number of short talks given by mid-career researchers, whereas at WS16 the ``Speaker'' category was reserved for keynote speakers.

% \begin{figure}
%   \centering
%   \subfloat[\textbf{WS16}\label{fig:s_groups_full_WS16}]{
%     \includegraphics[width=0.95\columnwidth]{s_groupes_full_WS16}
%   }
%   \\
%   \subfloat[\textbf{ICCSS17}\label{fig:s_groups_full_ICCSS17}]{
%     \includegraphics[width=0.95\columnwidth]{s_groupes_full_ICCSS17}
%   }
%   \caption{\textbf{Average individual interaction duration in each group.} For each group we compute the average interaction duration of the individuals, for the whole conference. The durations are measured in seconds. The error bar shows the standard error on the measure of this average strength. The solid colored line and colored regions show the overall average value for all nodes, with one and two standard deviations.}
%   \label{fig:s_groups}
% \end{figure}

The same effects are found when comparing the average interaction duration between groups, \emph{i.e.} the average time spent by each individual with each person he/she has been in contact with (Fig.~\ref{fig:groups}, middle row). Similarly to degree, most groups fall within the global average, while a few exhibit either significantly lower or higher interactivity. Other interactivity outliers are not necessarily the same as for connectivity and represent effects that are specific to the two contexts.
%, this is true for Language 2 (high) and Other Academic status (low) in WS16, and for Country 3 (high), Social/Political/Behavioural Scientists (low), Poster Presenters (high) and Staff (low) for ICCSS17. Other Academic Status and Staff can be understood by the same argument as precedently, Country 3 and Language 2 by cultural preferences. The fact that Social Scientists interacted less and Poster Presenters more during the ICCSS17 remains unexplained \textcolor{blue}{(we had two poster sessions at ICCSS17=more active time for poster presenters)}. Most importantly, all these results (except Other Academic Status) are specific to one conference only.
The only group difference that emerges identically and with clarity at both conferences is the lower connectivity, lower interactivity pattern of professors and participants above age 40 years. The consistency of this pattern suggests that a typical behavioural mechanism might be at play during academic gatherings.

% \begin{figure}
%   \centering
%   \subfloat[\textbf{WS16}\label{fig:T_groups_WS16}]{
%     \includegraphics[width=0.95\columnwidth]{T_groupes_WS16}
%   }
%   \\
%   \subfloat[\textbf{ICCSS17}\label{fig:T_groups_ICCSS17}]{
%     \includegraphics[width=0.95\columnwidth]{T_groupes_ICCSS17}
%   }
%   \caption{\textbf{Average daily individual presence duration in each group.} For each group we compute the average daily presence duration of the individuals for the whole conference. The durations are measured in hours. The error bar shows the standard error on the measure of this average strength. The solid colored line and colored regions show the overall average value for all nodes, with one and two standard deviations.}
%   \label{fig:T_groups}
% \end{figure}

The lower connectivity/interactivity of professors might simply be explained by the fact that professors spend only little time at the conferences, for example, they may attend only to give their talk and then leave again immediately after their session is finished. However, our data rule out this explanation. When looking at the average daily duration presence for each group (defined as the duration between the first and the last point in time where an individual has been in contact during each day) (Fig.~\ref{fig:groups}, bottom row), we see that most groups, including professors/age above 40 fall within the average global behaviour. The previous clich\'e does not hold: We find that professors connect and interact less than average, while spending as much time at the conferences as the other participants.

This result is important, as it shows unambiguously that the status-based attractiveness hypothesis does not hold in this context. The purpose of a scientific conference, the career-stage-specific needs and ways of creating capital in academia may explain this phenomenon. In terms of career-building and expertise-building, conferences are sites for the accumulation of academic capital, which encompasses building both relations and expertise. While we can safely assume that researchers accumulate such capital throughout their careers, they may do so more actively during some career stages than during others. The data we obtained suggest exactly that: Mid-career researchers are the ones who (have to) actively work on their assets, whereas professors can capitalise on the social capital or resources they already own. Conferences are an opportunity for researchers to present and exchange ideas, but also to become known and explore new career opportunities through \emph{networking}. Non-tenured, early-career researchers are indeed expected to interact, demonstrate their expertise and promote themselves actively, as they are those who need this networking to advance their careers. Professors, as established researchers, who are high in status and accumulated expertise may not \textit{need} to network as actively (anymore). Based on these premises, we can reinterpret our results: It is not professors who show a lower connectivity/interactivity. Rather, it is all the non-tenured, early-career researchers who connect/interact more. These results thus indicate that conferences are used as a tool for self-promotion, which most strongly concerns groups in precarious positions. As a final note, let us remark that the difference is not between interacting and not interacting, only that there is a significant difference in the intensity of interactions. This raises the question of the interplay between quantity and quality of interaction. One could interpret the results as established researchers such as tenured professors interacting more efficiently or, on the contrary, only engaging in socialising, \emph{i.e.} more superficial interactions, while non-tenured would actively engage in academic discussions. It can also be argued that professors have already accumulated expertise in their field, built their concepts and contributed major work while post-docs are in a phase of accretion. This argument works both for ``knowledge'' and ``status'' --- which makes for a more general line of explanation of the existence of an ``accumulation phase'' vs. a ``saturation phase''.

The differences we observe could be associated with different endpoints of communicative interactions in both groups: More strategic and task-driven ones on the side of non-tenured researchers, more relational (``social'') and ``shared-interest''-driven ones on the part of professors. The findings are also in line with some results from scientometric studies that found collaboration curves to show an inverted U-shape, with the highest incidence of both disciplinary and interdisciplinary collaborative action for mid-career-researchers \cite{vanRijnsoever:collaboration}. However, \cite{milojevic2012academicage} showed that when it comes to citing behaviour or productivity, there are no significant differences related to academic seniority, and \cite{zuckerman1972age} showed a correlation between (academic) age and the receipt of citations. What appears to be contradictory at first sight is no longer so contradictory if we differentiate types of academic action (networking, publishing productivity, citing, being cited) and allocate them to different phases. Intellectual output (publishing) and the processing of ideas (citing) is important throughout an academic career.Whereas mid-career researchers are more active in exploring collaborations and weaving intellectual networks to build their capital, the solid position of established researchers is sustained by their long-established contacts and ascribed capital, on the basis of which they receive even more capital (e.g. citations).

\subsection{Mixing, homophily and avoidance behaviours}

\begin{figure*}
  \centering
  \subfloat[\textbf{WS16}\label{fig:CM_Discipline_WS16}]{
    \includegraphics[width=0.95\columnwidth]{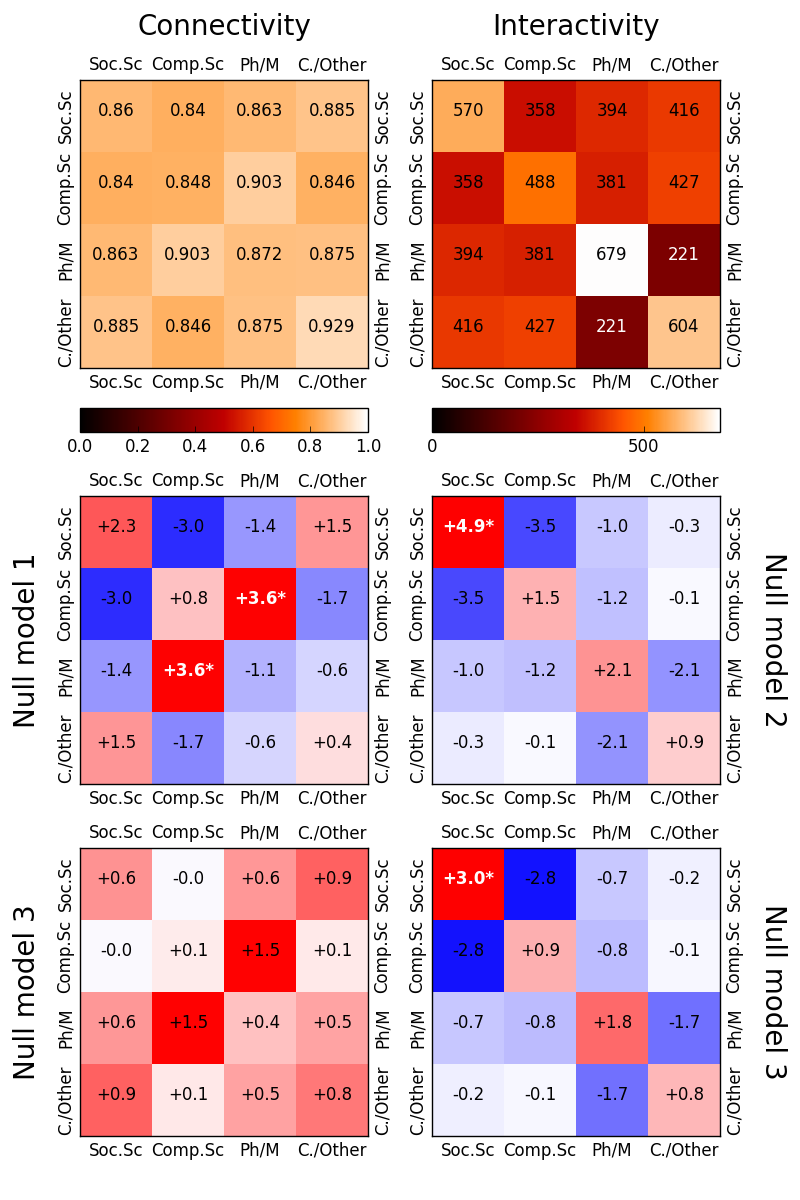}
  }
  \subfloat[\textbf{ICCSS17}\label{fig:CM_Discipline_ICCSS17}]{
    \includegraphics[width=0.95\columnwidth]{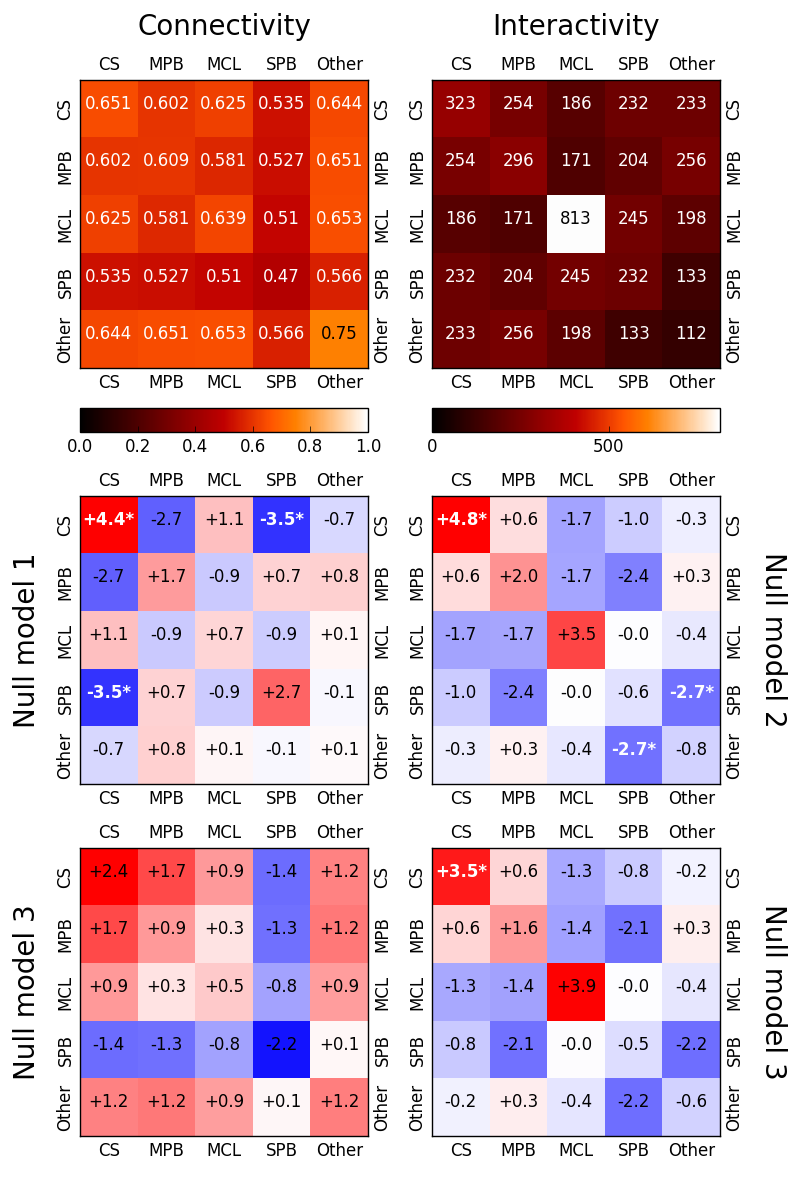}
  }
  \caption{\textbf{Contact matrices for disciplines.} Next, we assess how well participants from different disciplines mix. We compute for each conference the contact matrix in connectivity (\emph{i.e.} link density) and interactivity (\emph{i.e.} average total contact duration) considering the discipline (top row). We then test the statistical significance of the values of the contact matrices by performing three different null models: \Pk (Null model 1), \Ppw (Null model 2) and \Piso (Null model 3). Deviations are indicated in number of standard deviations $\sigma$ from the mean (\emph{i.e.} z-score). Positive deviations are in red and negative deviations in blue. Deviations marked in white with a star are significant under $p < 0.01$ (taking into account a Bonferroni correction). While some elements are significant outliers, there is no consistency across null models or conferences. Furthermore, these outliers are rare, which indicates that disciplines mix rather well.}
  \label{fig:CM_Discipline}
\end{figure*}

\begin{figure*}
  \centering
  \subfloat[\textbf{WS16}\label{fig:CM_Gender_WS16}]{
    \includegraphics[width=0.95\columnwidth]{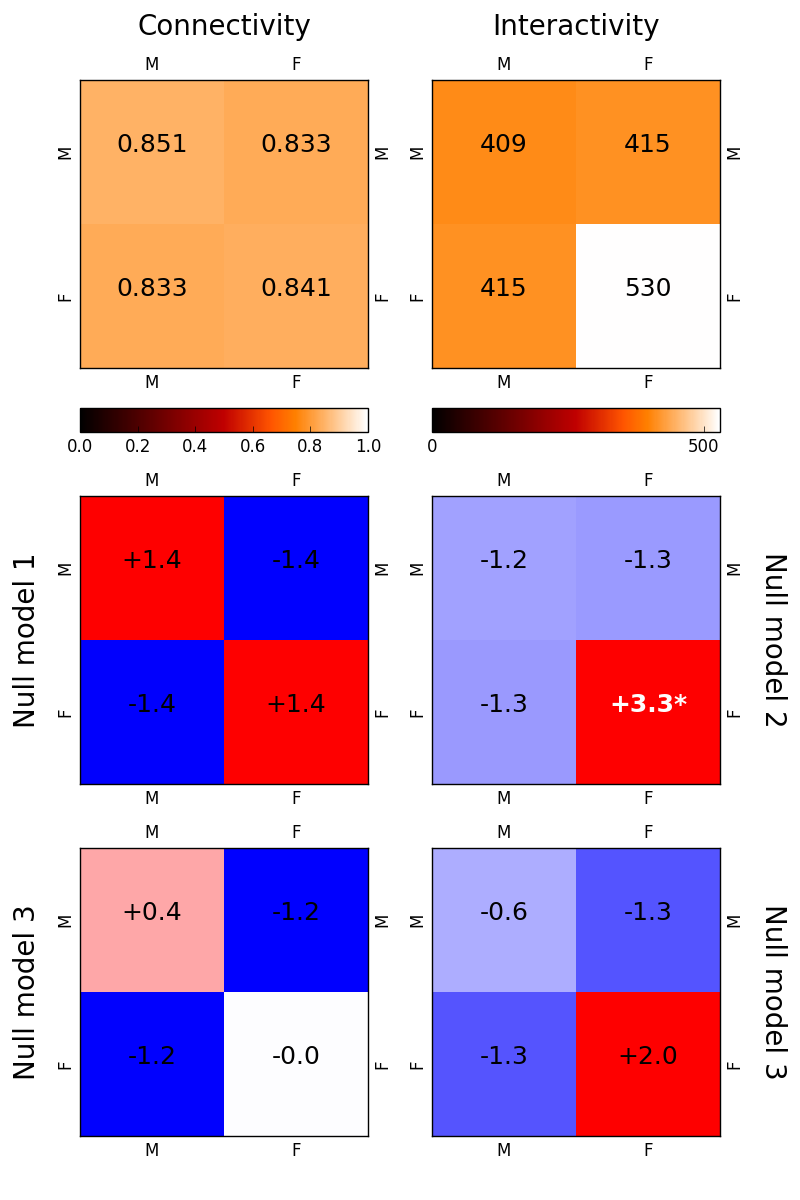}
  }
  \subfloat[\textbf{ICCSS17}\label{fig:CM_Gender_ICCSS17}]{
    \includegraphics[width=0.95\columnwidth]{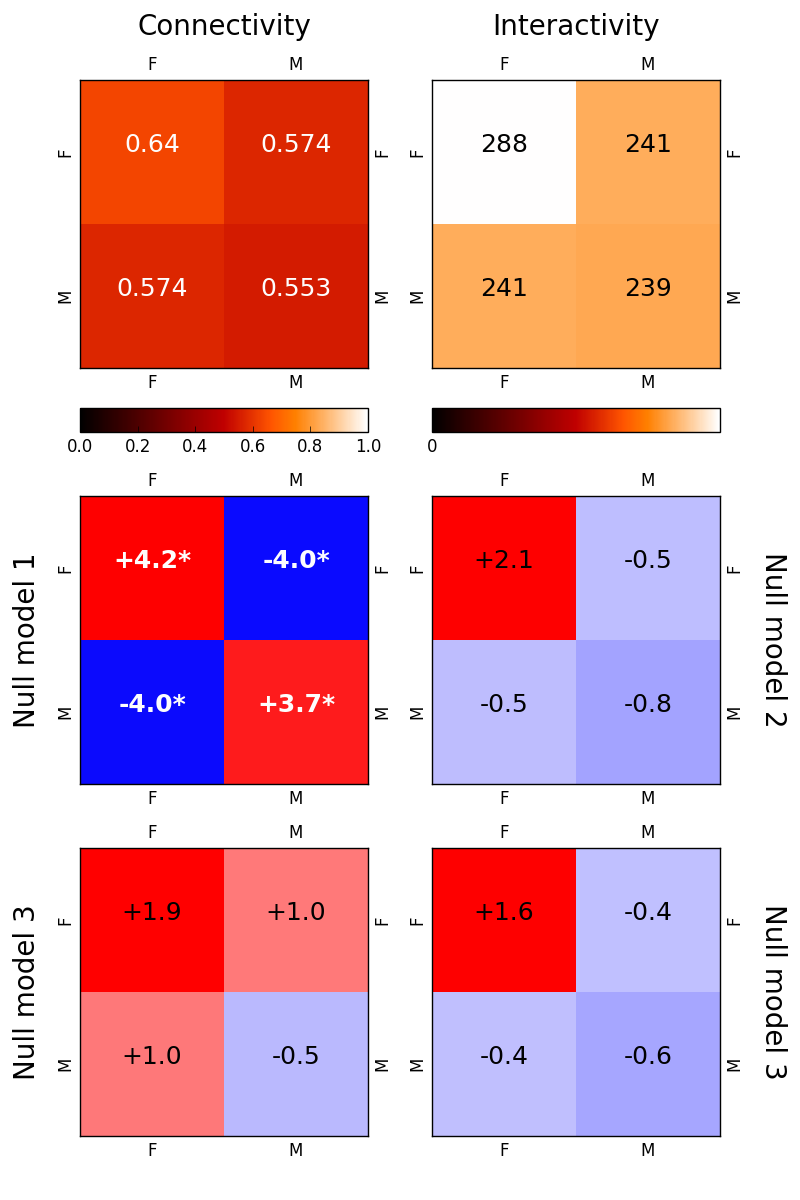}
  }
  \caption{\textbf{Contact matrices for genders.} We compute for each conference the contact matrix in connectivity (\emph{i.e.} link density) and interactivity (\emph{i.e.} average total contact duration) considering the gender (top row). We then test the statistical significance of the values of the contact matrices by performing three different null models: \Pk (Null model 1), \Ppw (Null model 2) and \Piso (Null model 3). Deviations are indicated in number of standard deviations $\sigma$ from the mean (\emph{i.e.} z-score). Positive deviations are in red and negative deviations in blue. Deviations marked in white with a star are significant under $p < 0.01$ (taking into account a Bonferroni correction). While some elements are significant outliers, there is no consistency across null models or conferences. Furthermore, these outliers are rare, which indicates that no particular pattern related to gender exists in these conferences.}
  \label{fig:CM_Gender}
\end{figure*}

\begin{figure*}
  \centering
  \subfloat[\textbf{WS16}\label{fig:CM_Language_WS16}]{
    \includegraphics[width=0.95\columnwidth]{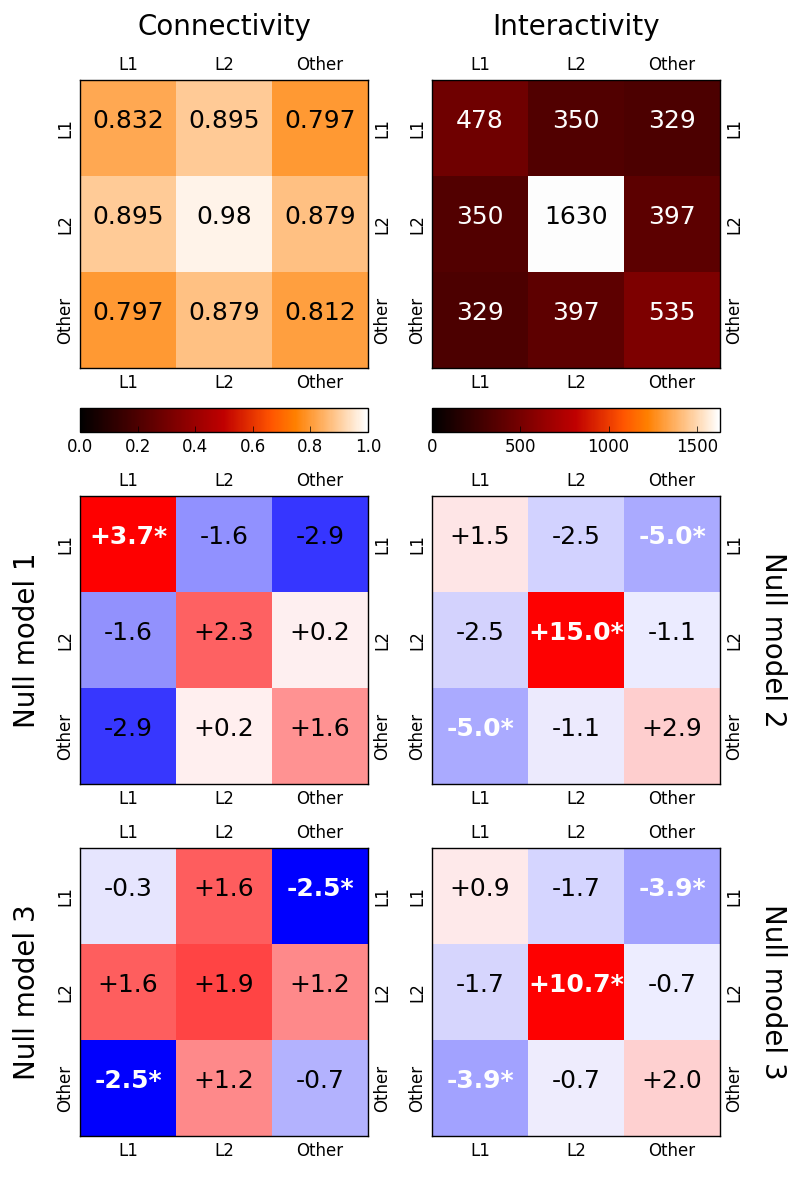}
  }
  \subfloat[\textbf{ICCSS17}\label{fig:CM_Language_ICCSS17}]{
    \includegraphics[width=0.95\columnwidth]{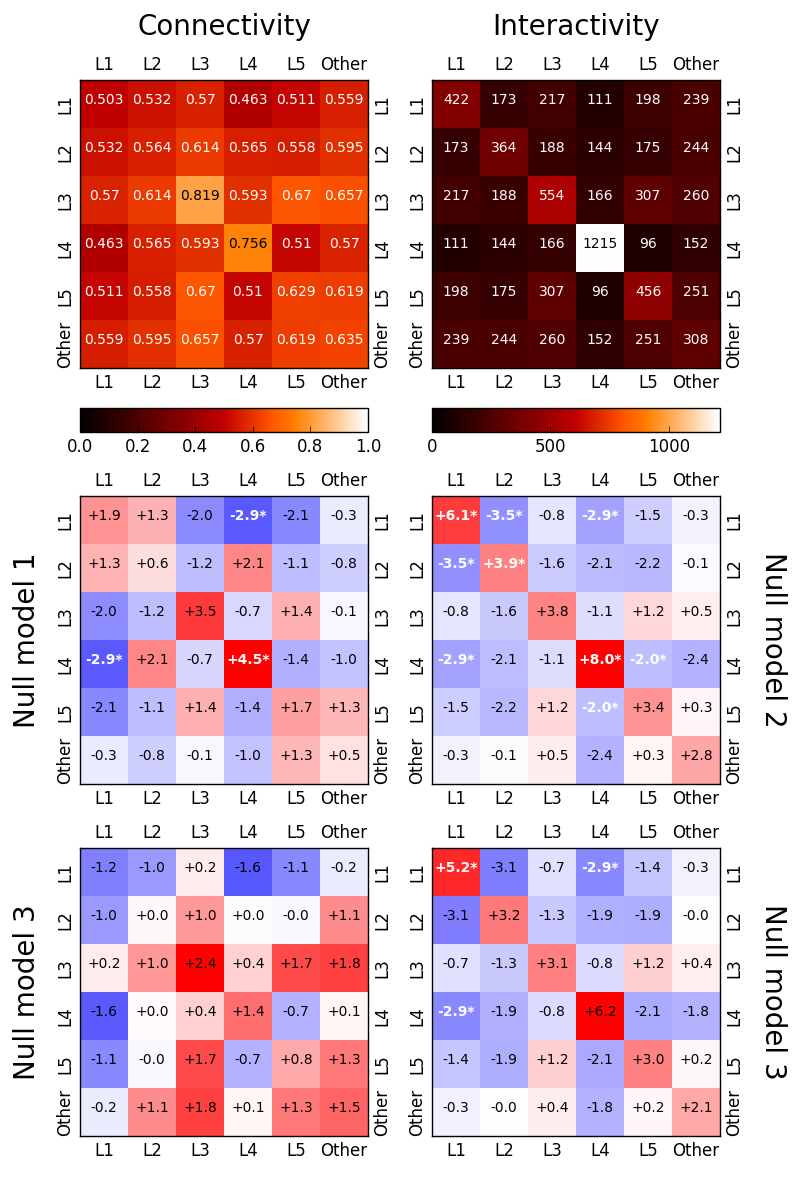}
  }
  \caption{\textbf{Contact matrices for languages.} We compute for each conference the contact matrix in connectivity (\emph{i.e.} link density) and interactivity (\emph{i.e.} average total contact duration) considering participants' language (top row). We then test the statistical significance of the values of the contact matrices by performing three different null models: \Pk (Null model 1), \Ppw (Null model 2) and \Piso (Null model 3). Deviations are indicated in number of standard deviations $\sigma$ from the mean (\emph{i.e.} z-score). Positive deviations are in red and negative deviations in blue. Deviations marked in white with a star are significant under $p < 0.01$ (taking into account a Bonferroni correction). Except for the Null model 3 on connectivity, deviations are always positive on the diagonal of the contact matrices, with several being statistically significant, which indicates the presence of language-based homophily.}
  \label{fig:CM_Language}
\end{figure*}

For each attribute category, we compute the connectivity and interactivity contact matrix (see Methods for the mathematical definition of contact matrices). These matrices allow us to analyse connectivity and interactivity patterns for each pair of groups. The contact matrices for both conferences and each set of attributes and are reported in the SI (see Figure \ref{fig:CM_Discipline}, \ref{fig:CM_Gender} and \ref{fig:CM_Language} top rows for the cases of discipline, gender and language groups, respectively).

Both connectivity and interactivity appear fairly similar between group pairs. The exceptions are all associated with the outliers identified in the previous section. Most of these outliers are related to intra-group behaviour, particularly for connectivity and interactivity that are higher than average. This could sign homophilic behaviour, \emph{i.e.} the tendency to connect and interact with individuals from the same group. Negative outliers on the contrary can arise from avoidance strategies, where individuals from two groups connect and interact less than average.

We test whether these differences in connectivity and interactivity are evidence for preferential connection behaviour or avoidance effects using a null model framework. The procedure consists in randomising the network while retaining a few chosen properties, to generate a distribution of possible random outcomes. We then compare the empirical observation to the distribution, and test the statistical significance of each measure by computing z-scores. In the present case, we define 3 different baselines, to account for three possible effects (see Methods for the description of the null models). Figure \ref{fig:CM_Discipline}, \ref{fig:CM_Gender} and \ref{fig:CM_Language} show the cases of discipline, gender and language groups, respectively (see SI for the other attributes, along with a table listing all significant deviations).

Overall, the results are the following:
\begin{itemize}
\item Both conferences are rather well-mixed.
\item Almost all methods show no significant preferences or avoidance related to gender.
\item Participation to a previous edition shows no signal (see SI).
\item There are a few specific cases of positive bias within disciplines and negative bias between disciplines. However, these behaviours are not consistent across conferences, and usually do not represent the majority (Fig.~\ref{fig:CM_Discipline}).
\item Null model 3 shows a clear separation in behaviour between tenured and non-tenured researchers in connectivity (see SI). Tenured researchers connect significantly less, both internally and with other groups, while non-tenured connect more among themselves. This effect is present considering either academic status, age group or role at the conference. Tenured researchers are indeed usually professors, above age 40 and more likely to be speakers, whereas non-tenured researchers are PhD students, postdoctoral research, or associate/assistant researchers, are under age 40 and more likely to be poster presenters.
\item A clear signal of country/language-based homophily is present in the network, not necessarily significant but systematically present  (Fig.~\ref{fig:CM_Language}).
\item Staff usually constitute a group clearly separated from the crowd.
\end{itemize}

%As expected, we recover the behavioural patterns previously described, concerning how permanent and non-permanent researchers use conferences differently. The latter connect and interact with each other, focusing on the networking aspect of the event, while the first exhibit significantly lower connecting and interacting behaviour. More surprising, no signal for gender related behaviour appear, indicating that this attribute does not seem to have an effect on mixing. Finally, although Academia is intrinsically international, there appear to be clear signals of country/language preferential interactions.

For academic seniority (see SI), we find status homophily in the sense that mixing occurs horizontally rather than vertically, (\emph{i.e.} a peer conversation is more likely than a status-diverse-contact like a mentoring situation. This implies that the vertical, mentoring type of networking seems to be less prevalent than the horizontal, peer-oriented type of interaction.

With regard to the overall mixing behaviour (Fig.~\ref{fig:CM_Discipline}), we can argue that the CSS crowd appears to be highly interdisciplinary and that every sub-community is generally eager to get access to skills and expertise from (all) others. On a structural level, this corresponds to findings that researchers working in applied disciplines (i.e. disciplines directed towards practical applications) engage more in interdisciplinary collaboration than those in basic disciplines \cite{vanRijnsoever:collaboration}. This effect might be even stronger considering that CSS can still be seen as an emerging research field with no well-established disciplinary identity but benefiting from a plenitude of research interests in ``the digital social'' from various disciplines. As a consequence, networking and research collaboration in CSS is inclined to be interdisciplinary. Yet there are certain affinities which seem to be stronger than others: For example, physicists interact strongly with both social scientists and computer scientist, whereas the two latter groups interact below average with each other --- which puts physicists in a kind of brokerage position. It might be interesting to further observe these formative processes in order to find out about overlaps and interactions, brokerage positions or, eventually, get hold of the formation of a distinct disciplinary ``CSS-core''.

As for internationality/cosmopolitanism (Fig.~\ref{fig:CM_Language}), participants mixed well in terms of language (i.e. mother tongue) and current country of residence: English obviously serves as the lingua franca in the research field. Yet, we observed language/country homophily that might point to linguistic and geographical proximity as factors that facilitate networking and the formation of collaborative ties. Another explanation could be that researchers from a given country are more likely to know each other in person before the conference. Such pre-existing relationships would then translate into a higher likelihood of interacting at the conference. The signals of language/country-based homophilies are not equivalent; some groups appear more susceptible than others to connect and/or interact internally. In those cases, homophily might be an effect of a number of shared attributes in that group (and thus be a cumulative effect rather than a purely linguistic one); it can also be that language-related homophilic pattern we observed might have been fostered by the (large) size of the language group --- with a large interdisciplinary pool like that, there might be less need to transgress language borders in order to access the expertise a researcher needs. If this is true it would be a signal for internationality not being pursued as a value in itself.

Finally, we do not find any consistent gender-based homophily or heterophily in the data (Fig.~\ref{fig:CM_Gender}). Women represented a minority in both cases (39\,\% at WS16, 31\,\% at ICCSS17), but the only significant homophily signal appears for connectivity in ICCSS17, with respect to the Null model 1.

\subsection{Link dynamics}\label{sec:flow}

\subsubsection{Link flows}

\begin{figure}
  \centering
  \subfloat[\textbf{Link flows --- WS16}\label{fig:flows_WS16}]{
    \includegraphics[height=0.29\columnwidth]{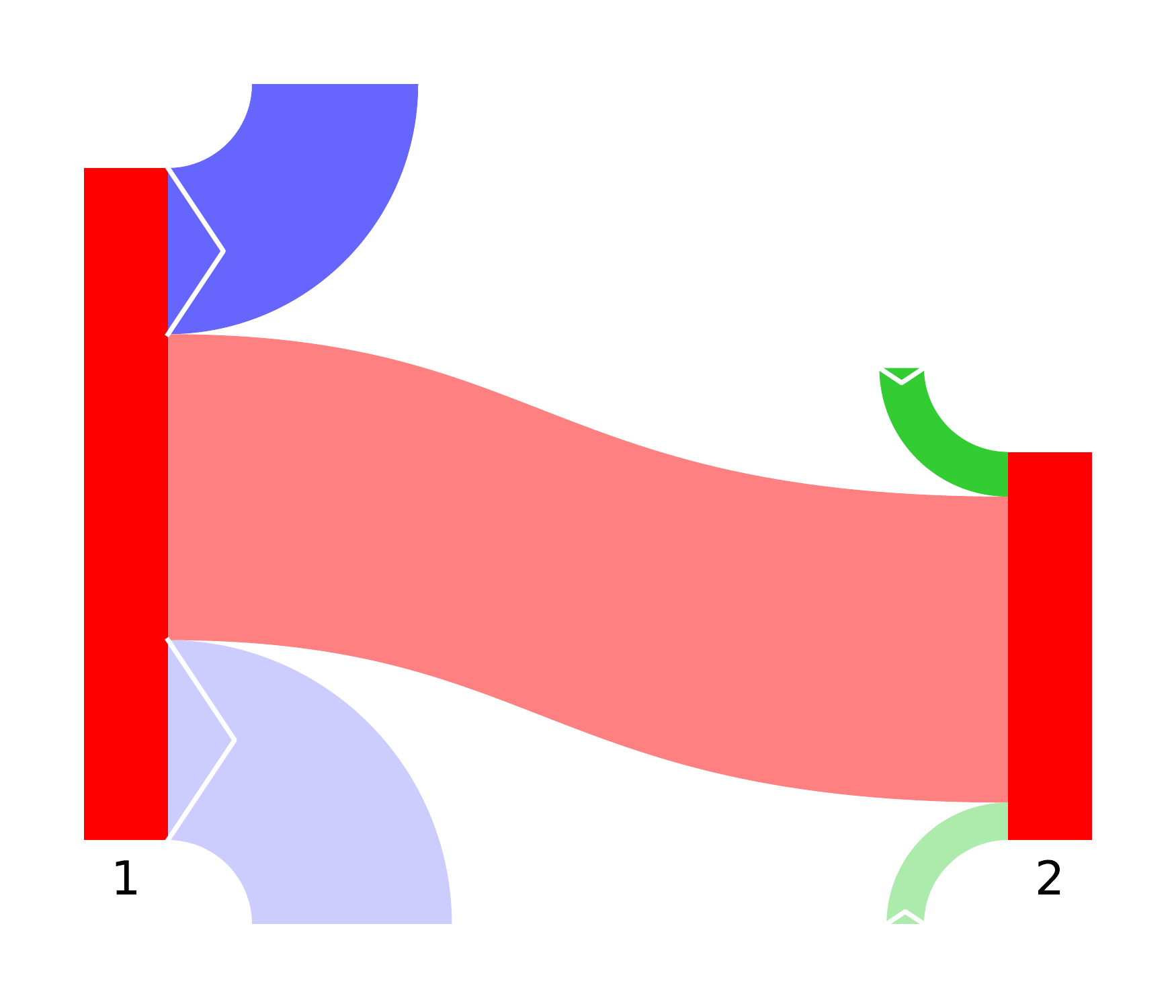}
  }
  \subfloat[\textbf{Link flows --- ICCSS17}\label{fig:flows_ICCSS17}]{
    \includegraphics[height=0.29\columnwidth]{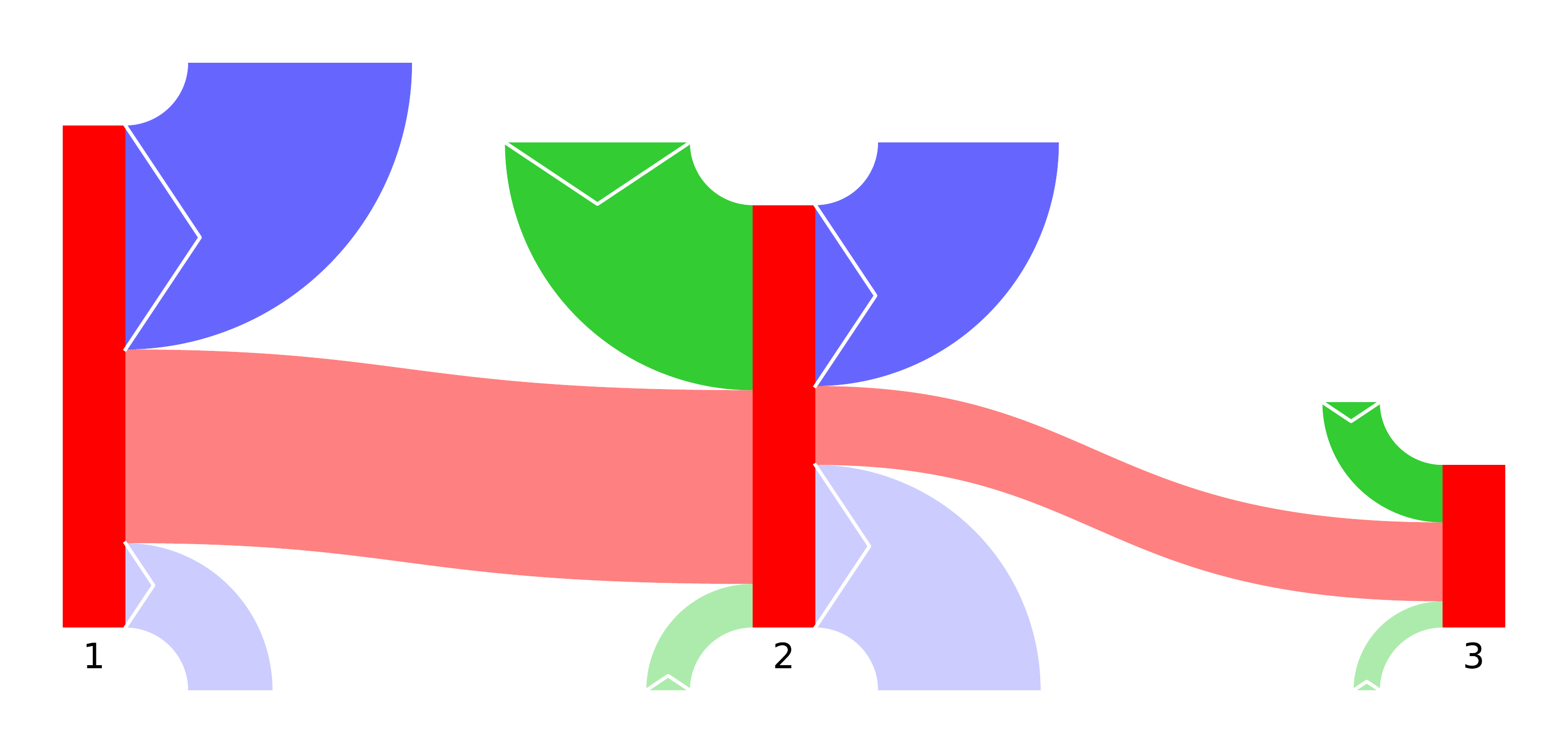}
  }
  \\
  \subfloat[\textbf{Link loss --- WS16}\label{fig:link_loss_WS16}]{
    \includegraphics[height=0.31\columnwidth]{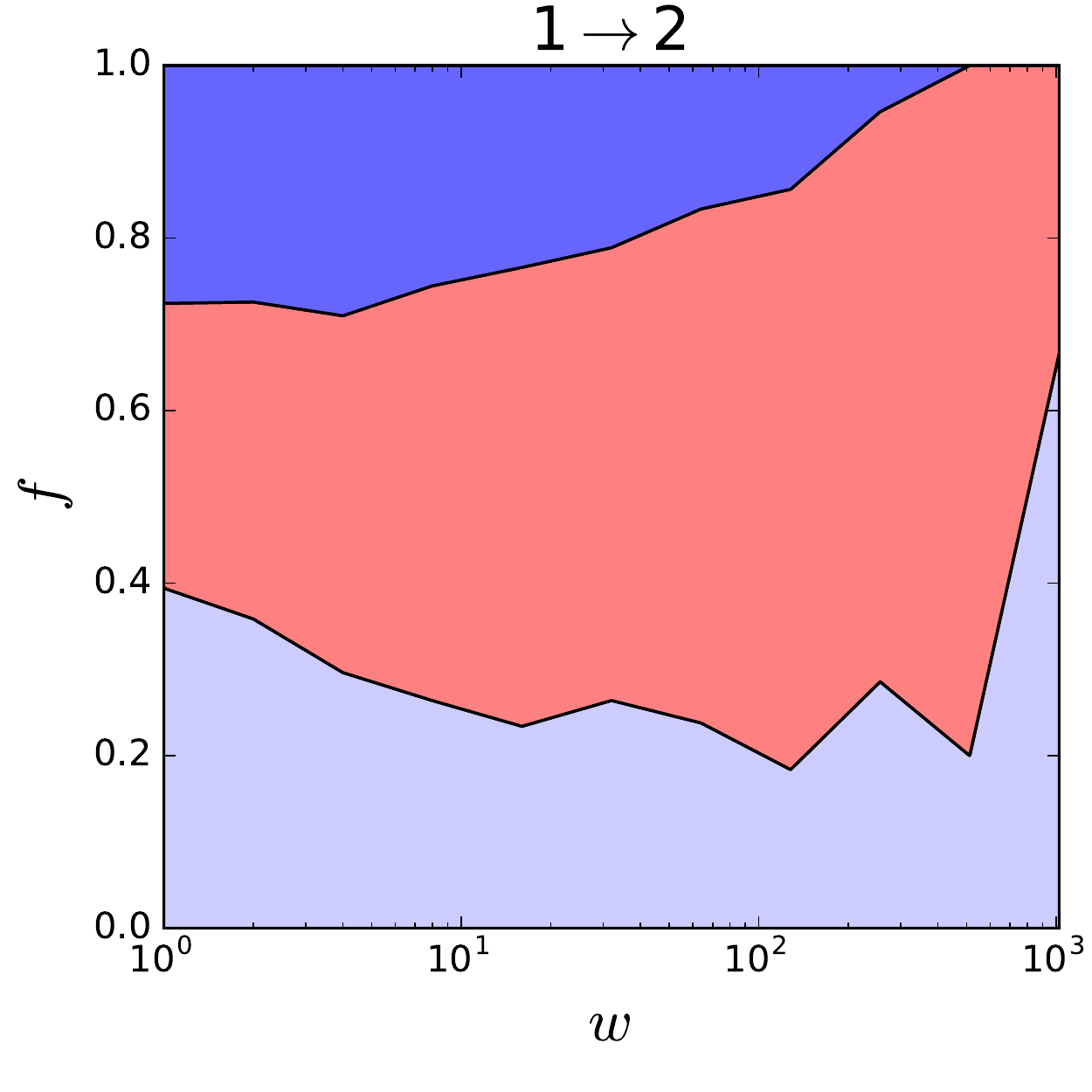}
  }
  \subfloat[\textbf{Link loss --- ICCSS17}\label{fig:link_loss_ICCSS17}]{
    \includegraphics[height=0.31\columnwidth]{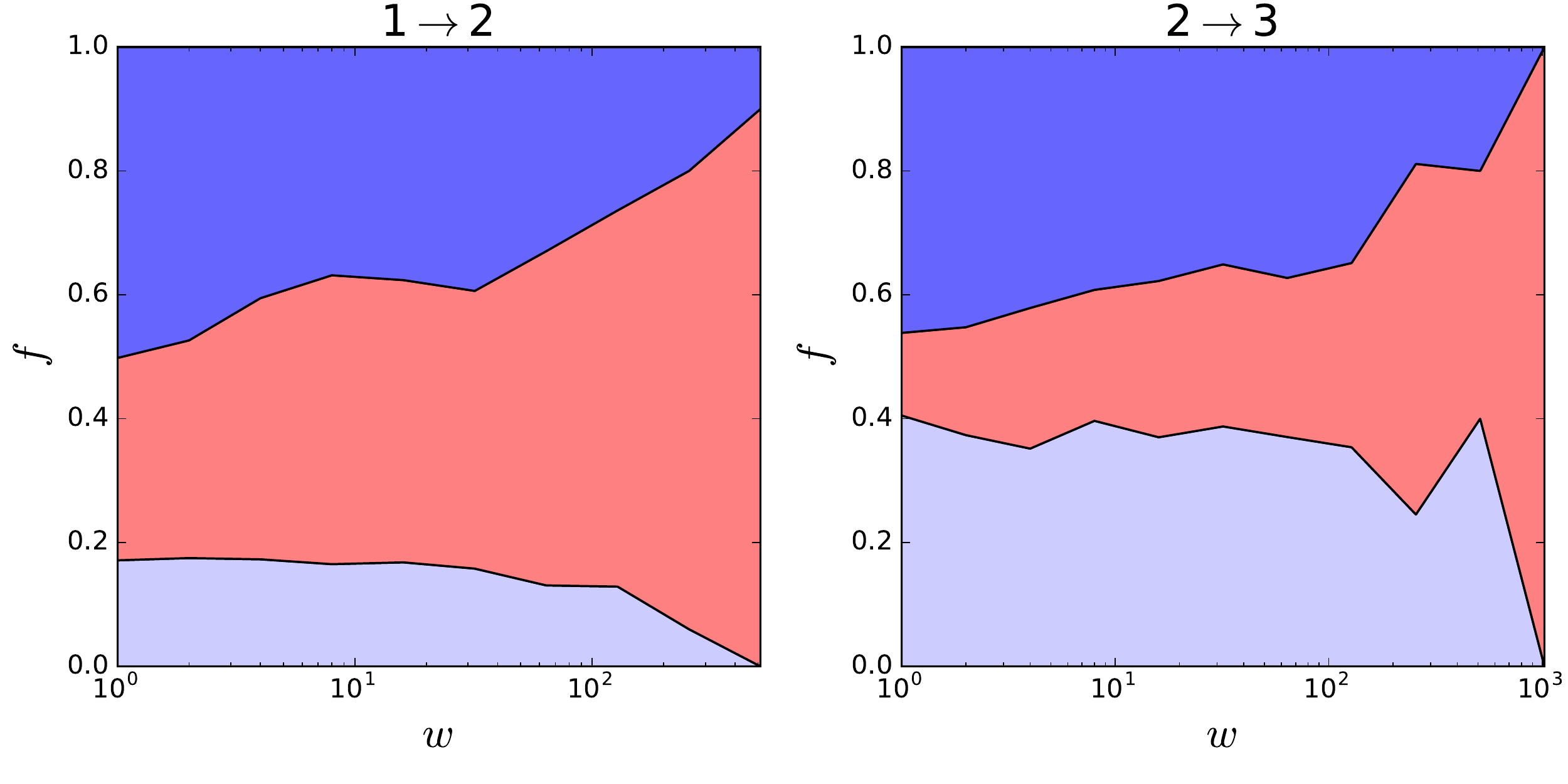}
  }
  \\
  \subfloat[\textbf{Link gain --- WS16}\label{fig:link_gain_WS16}]{
    \includegraphics[height=0.31\columnwidth]{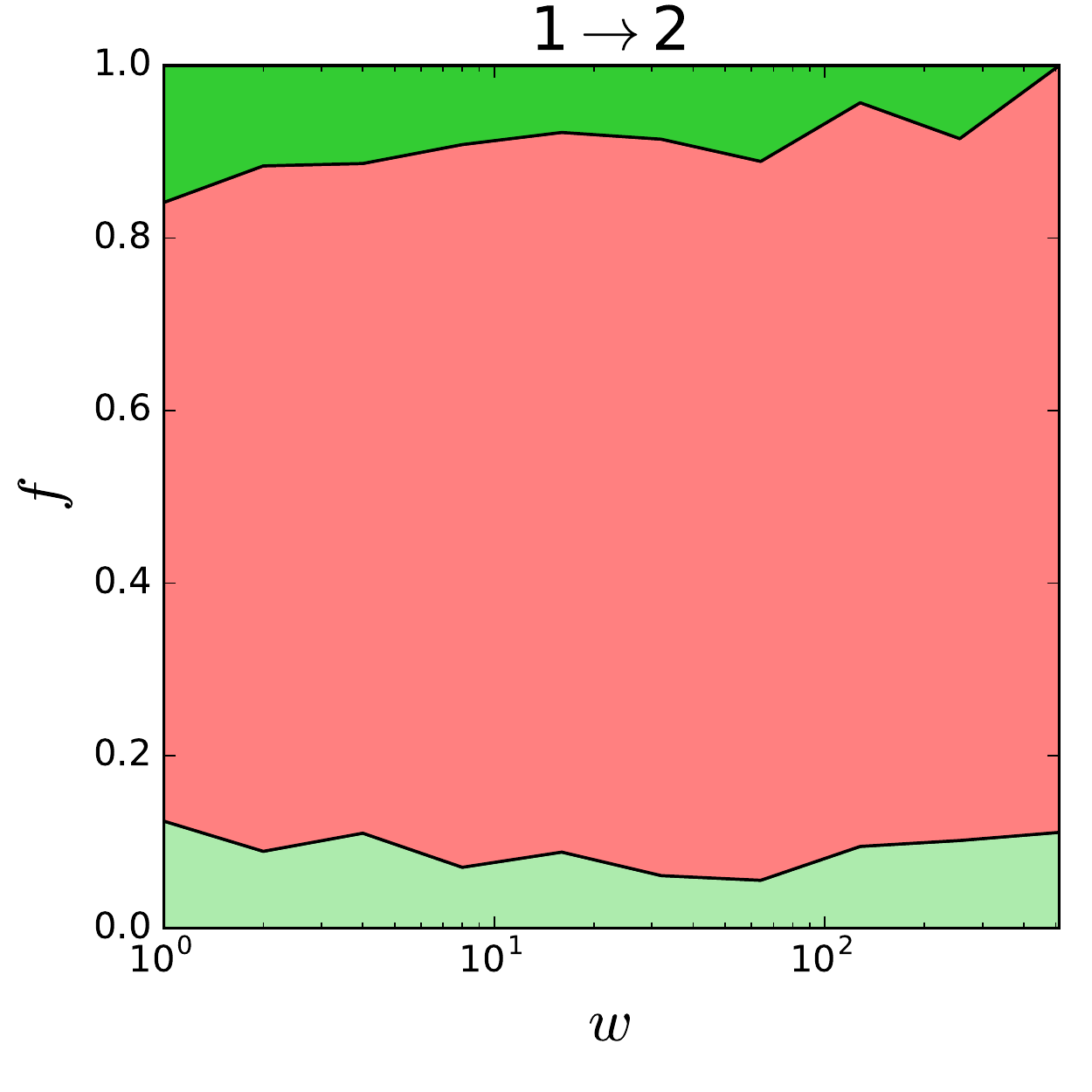}
  }
  \subfloat[\textbf{Link gain --- ICCSS17}\label{fig:link_gain_ICCSS17}]{
    \includegraphics[height=0.31\columnwidth]{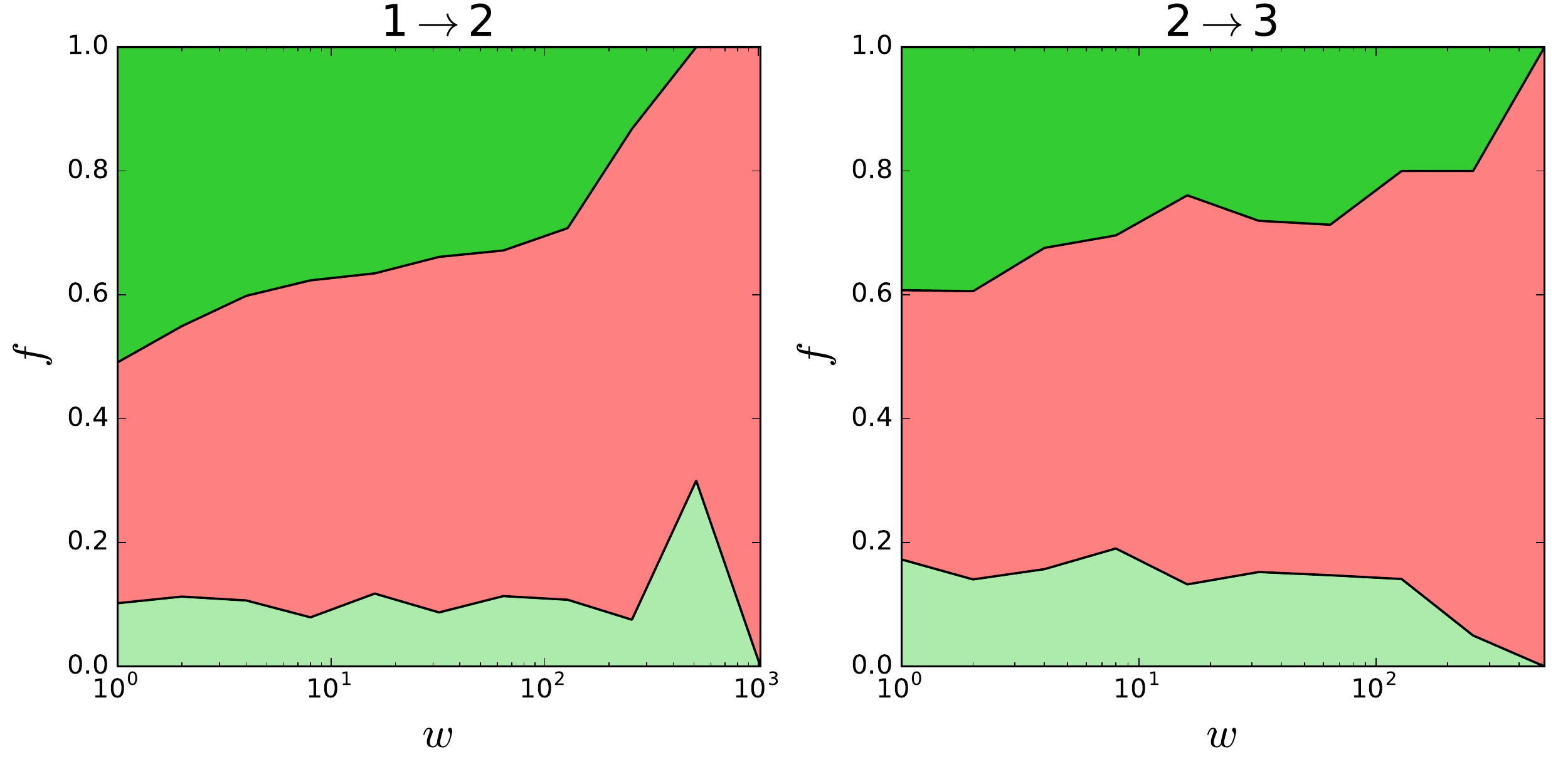}
  }
%   \\
%   \subfloat[\textbf{Values}\label{tab:flows}]{
%     \begin{tabular}{|c||c|c||c|c|c|}
%       \hline
%       & \multicolumn{2}{c||}{\textbf{WS16}} & \multicolumn{3}{c|}{\textbf{ICCSS17}} \\
%       \hline
%       Day       &    1 &    2 &    1  &    2 &    3 \\
%       \hline
%       Edges     & 6681 & 3858 & 11125 & 9359 & 3602 \\
%       Node loss & 1987 &    0 &  1870 & 3603 &    0 \\
%       Node gain &    0 &  372 &     0 &  966 &  581 \\
%       Link loss & 1652 &    0 &  4963 & 4008 &    0 \\
%       Link gain &    0 &  444 &     0 & 4101 & 1273 \\
%       Conserved & 3042 &    0 &  4292 & 1748 &    0 \\
%       \hline
%     \end{tabular}
%   }
  \caption{\textbf{Link dynamics between days.} \textit{Link flows} (a and b): For each pair of days, we compute the total amount of links (red bars), the amounts of lost links (blue arrows) separating the loss due to nodes exiting the system (light blue) and links that disappeared between nodes that were present in both days (dark blue), the amounts of links gained (green arrows) separating gain due to entering nodes (light green) and new links appearing between nodes that were present in both days (dark green), and the amounts of links conserved between the two days (red band). Numbers label the days. Table reporting the numerical values can be found in the SI. --- \textit{Distribution of links per bin of weight} (c to f): We compute for each pair of days the proportion of links belonging to each flow, per bin of weight in a log scale. We consider weights of the first day of the pair for link losses and weights of the second day of the pair for link gains. The colour code is the same as for flows. It shows that loss and gain due to node flows (pale blue and green) are mostly independent of weight, while the probability for a link to be conserved increases with its weight.}
  \label{fig:dyn}
\end{figure}

Both conferences exhibit a drop in the average degree as the event unfolds (see Figure \ref{fig:flows_WS16} and \ref{fig:flows_ICCSS17}, and the daily comparisons of degrees between groups in the SI). We are interested in understanding more precisely what happens to the interactions between the participants as time passes. To do so, we compute the evolution of the daily aggregated network from one day to the next. We count how many links are lost, gained and conserved between each pair of days. As the population is not fixed, we separate loss and gain in two parts: link dynamics within the stable part population (\emph{i.e.} nodes that are present on both days), and evolution due to the exiting or the entering of nodes in the system. The results for both conferences are reported in Fig.~\ref{fig:dyn}.

It is immediately clear that there is a large turnover in links from one day to the next. At both conferences, less than half of the links established during the first day are conserved on the second day (46\,\% for WS16, 39\,\% for ICCSS17). For the ICCSS17 case, only 19\,\% of the links from Day 2 are conserved on Day 3. Link losses and gains are more or less equally distributed between node-related and internal flows for WS16, while for ICCSS17 it is only true for link losses on Day 2. For the rest, internal turnover is much more important than node-related turnover. 

We decompose these flows according to the weights of the links (Fig.~\ref{fig:dyn}c to f). Although the signal is not completely clear for WS16 due to a relatively small number of links, it still appears that links that are conserved tend to have higher weights, while links that are lost or gained due to internal turnover tend to have lower weights. However, this separation is not complete: low-weight links can be conserved, high weight links can be gained and lost. Nonetheless, this indicates that a filtering on the links according to their weight is at play on this internal turnover. On the contrary, turnover due to node exiting and entering the system appears to be uncorrelated with the weight of the link, as the fraction is the same whichever the weight (with some fluctuations for high weights, due to small number effects).

\subsubsection{Analysis of the conserved links}

\begin{figure}
  \centering
  \subfloat[\textbf{WS16}\label{fig:evol_d_WS16}]{
    \includegraphics[height=0.32\columnwidth]{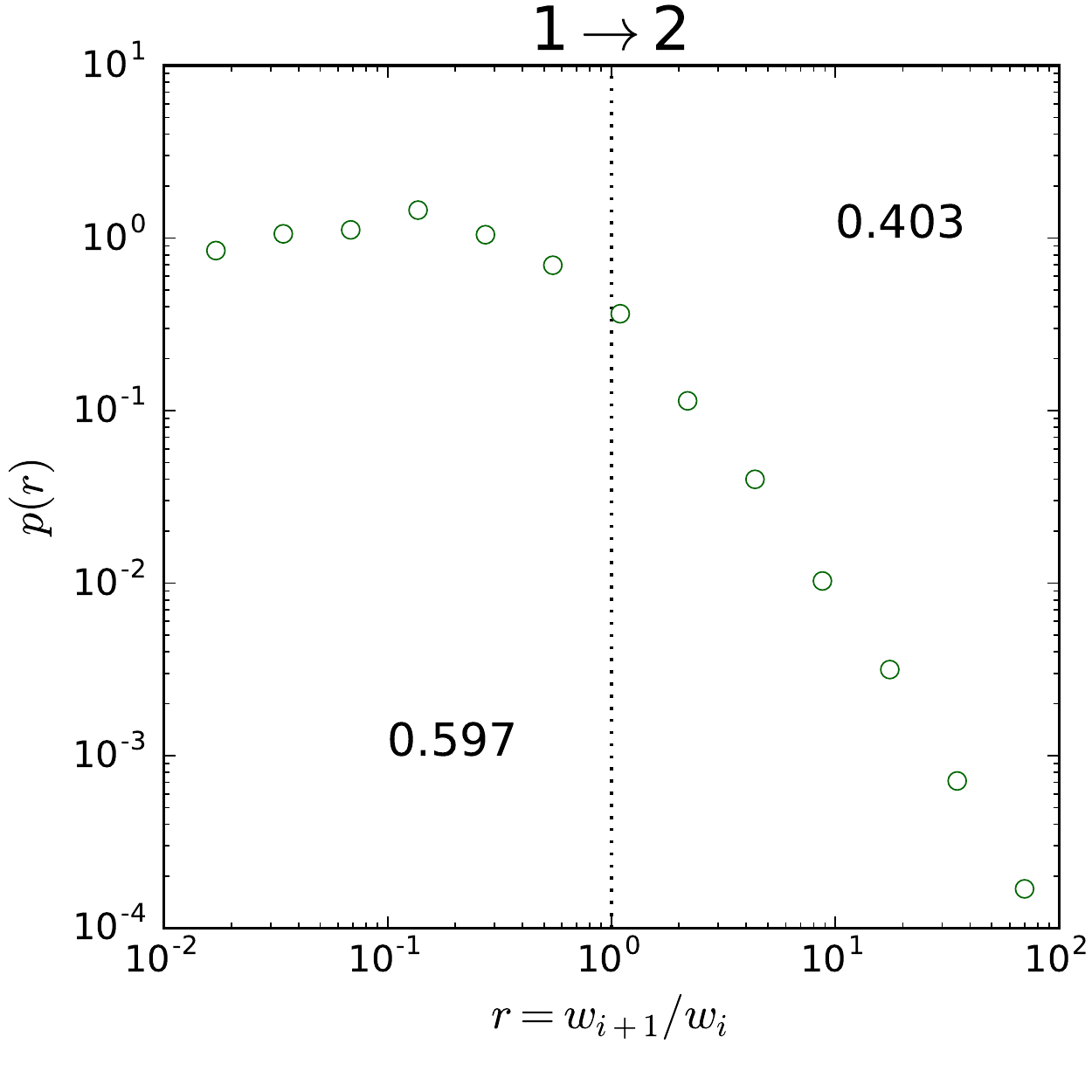}
  }
  \subfloat[\textbf{ICCSS17}\label{fig:evol_d_ICCSS17}]{
    \includegraphics[height=0.32\columnwidth]{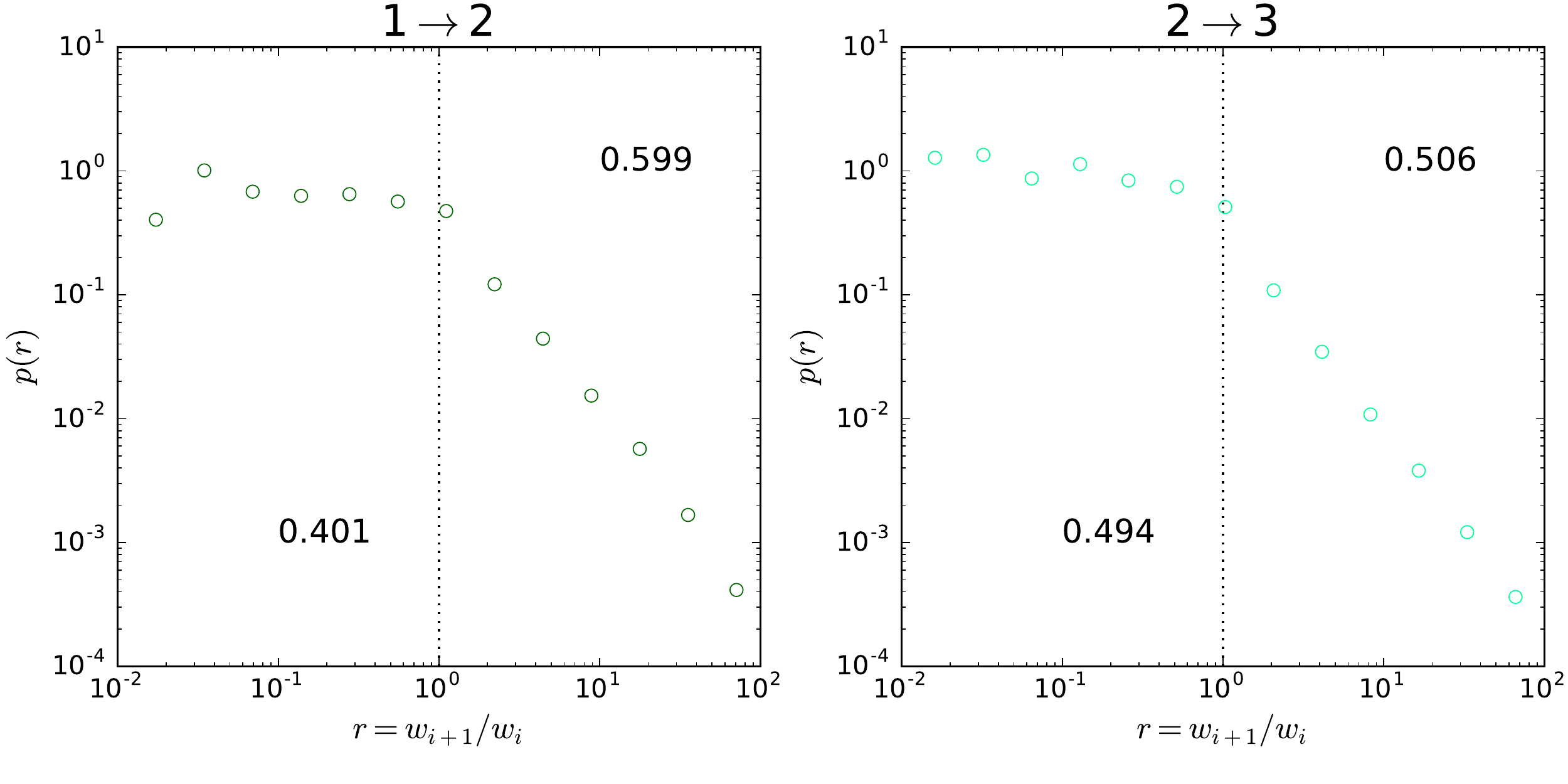}
  }
  \caption{\textbf{Evolution of the link weight for conserved links.} For each link that is conserved between two successive days, we compute the ratio $r$ between the total duration of the interaction $w$ (weight of the link) on day $i+1$ and on day $i$ ($r = w_{i+1}/w_i$). The dotted line marks $r=1$, which separates interactions that are stronger on the second day (right) and interactions that are weaker on the second day (left). Number show the fraction of links on each side of the distribution.}
  \label{fig:evol_d}
\end{figure}

We focus on the links that are conserved from one day to the next. We compute the evolution of the weights of these links between the two consecutive days by simply looking at the ratios $r$ of the link weights ($r = w_{i+1}/w_i$). We plot the distribution of these ratios for each pair of consecutive days.

As seen on Fig.~\ref{fig:evol_d}, roughly half of the conserved links are reinforced and half weakened (40/60 for WS16, 60/40 and 50/50 for ICCSS17). This indicates that, on top of the filtering previously described where links are simply deleted, another filtering is at play, keeping only half of the remaining ones with an equal or stronger intensity.

Second, the behaviour appears different for weakening and reinforcement. For $r<1$, the distribution is more or less flat, indicating that weakening occur with roughly equivalent probabilities. On the contrary, the probability to have values of $r>1$, \emph{i.e.} to see a large reinforcement of the interaction, decreases as $r$ increases. This reinforces the idea of a filtering: if a connection is marked as ``discardable'', then there is no incentive to enforce any particular weakening effect. However, reinforcing a connection implies an effort, the more intense the higher the effect, and therefore it is less likely to get high values of $r$.

The filtering down of contacts over the course of a conference could be motivated by the pursuit of efficacy and might constitute a signal of functional selection at work. This appears to be convincing especially with regard to the ``discarded'' links and the concentration on few links only. It is tempting to assume that the maintenance of weak links on days 2 or 3 might serve relational/social purposes and only the few intensified contacts point to a strong networking interest. Contrarily, the large impact of link turnover indicates that even in a highly functional context human communication does not appear to be strictly efficient or, to turn the argument, that exuberance, new influx and choice are important for social interaction which has to be kept open and multipurposeful. This is furthermore supported by the fact that the distributions of $r$ all adopt the same shape whichever the days of the conference. This is perhaps the most surprising, and might be the mark of an underlying, general rule about human face-to-face interactions.

When we compute the contact matrices in connectivity and interactivity considering only the socially filtered interactions, \emph{i.e.} the links that are both conserved and reinforced between two consecutive days (see SI). Qualitatively, the contact matrices are essentially very similar to the global ones. This indicates that the filtering process does not follow any particular pattern related to socio-demographic attributes, but rather occurs uniformly across the network. In the end, we may conclude that both the downsizing indolent towards attributes and the large turnover from day to day suggest that there is a strong situational impact on interaction, and that making one's network more specific means to reduce quantity rather than scope.

%In combination with the fact that we observe a downsizing in numbers but no changes in attributes of connections (see SI for the daily contact matrices) this finding may twist the functional perspective in an interesting way: whereas the distribution of socio-demographic, disciplinary and status related features remain stable and apparently are deliberate from the start the actual choice of individuals you communicate with develops seems to be situational and to some extent arbitrary

\section{Discussion}

%\hl{Rather than a discussion (of the results), which you have done in the results section, I would write some conclusions (how results can be generalized, what can be learned, etc). I would like to learn about shortages of the analysis, particularly that you don't know who knew whom before he came to the conference. Also, address the privacy and anonymization issues and why you can't release all data (maybe arrange for the data to be accessible through the Secure Data Center?).}

We looked at patterns of scholarly interaction at two computational social science conferences with regard to the mixing behaviour of disciplinary, cultural, and age groups and the temporal dynamics of the network. Academic networking is at the heart of international conferences and especially interdisciplinary exchange is crucial for scientific innovation and the formation of new fields of research which emerge across established disciplines \cite{mertonsociologyofscience, kuhnstructure}. As much as novel ideas spring off non-similarity, contacts have to be enabled and made until they can connect diverse communities via overlaps. Such loosely intersecting cohesive groups generate shared knowledge and chains of conceptual agreement \cite{moody2004structure,friedkin2006structural}. Overall we see that the similarity attraction paradigm is not particularly prevalent at the CSS conferences we studied; it is rather diversity which is attractive: people intermingle well and strong containment tendencies cannot be observed for any segment, be it disciplinary, cultural or gender based. However, there are behavioural differences and barriers. With the combination of rich survey data on socio-demographics and sensor-based contact data, we can look into the granularity of such processes. We take these contacts as signals for the dissemination of discipline specific ideas and the mixing patterns as proxies for methodological trends.

Academic conferences are the places to get a position in more than one sense. With respect to academic seniority and role at the conference we find a clear distinction between younger scholars and established ones, with the first more actively accumulating relations and academic capital. This characterises different academic career phases with the need to obtain and display everything --- knowledge, contacts, visibility --- that may generate and sustain a strong position in the academic field and the exertion of such a position. This is in line with the Matthew effect \cite{merton1968matthew}, since the effect we observed can be interpreted as status execution taking less effort than status gain. The turnover in interactions can also be related to the idea that sociality is not absent in highly functional arenas; in human communication, strategic action cannot be purely instrumental and efficient --- there always have to be relational elements, redundancy and new, unexpected influx. Those ``social'' aspects of interaction appear to facilitate innovation, and so does the physical proximity of face-to-face-communication --- even in the digital age. In terms of practical implications, we can only encourage conference organisers to enable communication by providing breaks, networking slots and interactive formats such as poster sessions. We also advise participants who wish to benefit from those networking opportunities to attend academic events from the start.

The context of our study is rather narrow, as it focuses on international, interdisciplinary, scientific conferences which we admit are a peculiar setting. However, we believe that our method to study human interactions and social phenomena can be applied to a wide range of contexts. In particular, checking whether the relation we found between social capital and interactivity also appears in other situations, such as workplaces or institutions, would be very interesting. Our results are based on two distinct cases; however, we saw that there was already large variations in terms of behaviour between the two conferences, notwithstanding their numerous similarities. Obviously, our results would need to be checked with more studies. In particular, we assumed in our analysis that the population is in a ``blank state'' as the conferences start. This is far from true, as some people know each other beforehand. Such preexisting structure has definitely an impact on the behaviour during the conference. Controlling for this parameter would be another improvement in such a study. Nonetheless, the phenomena we have uncovered we presented represent a new step in the understanding of social mechanisms.

%ideas for generalization in terms of sociology of science:}
%\textcolor{blue}{\begin{itemize}
%\item{formation of discipline (CSS), interdisciplinarity}
%\item{conceptual agreement across groups (depends on loosely overlapping cohesive groups (cf. Moody, 2004; Friedkin, 1998)}
%\item{types of motivation for communication}
%\item{detect invisible colleges}
%\item{communication overlaps, shared knowledge generate 2chains of knowledge"}
%\item{general temporal pattern of link preservation}
%\item{"accretion"/"accumulation" vs. "saturation" (maybe putting it somewhat more politely)} 
%\item{why is it more functional to talk to peers ? easier, less risky, more rewarding?}
%\item{similarity attraction/ homophily vs. "barriers"}
%\item{conferences are the places to make a position}
%\item{collaboration as signal for the dissemination of discipline specific ideas (production of scientific knowledge)}
%\item {scientific fields emerge across established disciplines (Merton, Kuhn)}
%\item{mixing patterns of disciplines (as proxies for methodological trends)}
%\item{brokerage/ mediating/ central/ peripheral positions}
%\item{role of physical proximity, face-to-face-communication (even in digital age)}
%\end{itemize}}

\section{Methods}

\subsection{Data collection}

We used the SocioPatterns platform to collect face-to-face contacts between participants from two conferences. The setup of the RFID chips allows to detect physical proximity ($\sim$\,1.5\,m) when two individuals are standing in their respective front half-spheres (face-to-face). The temporal resolution is very high (20\,s) which allows for a very precise, \emph{in situ} and unbiased recording of these interactions. In addition to the SocioPatterns sensor data on face-to-face interactions, we gathered socio-demographic information about the participants, namely their age group, country of residence, mother tongue, academic status/seniority, disciplinary background, role in the conference and whether they attended one of the previous two editions of the conference.

The datasets were collected during the GESIS Winter Symposium on Computational Social Science, on November 30 and December 1, 2016 (WS16), and during the International Conference on Computational Social Science, July 10 to 13, 2017 (ICCSS17).

For WS16, there were 149 participants, from which 144 accepted to take part in the present study (96.6\,\%). Contact data was retrieved for 138 participants (92.6\,\%). Among them, we have at least partial socio-demographic information for 115 participants (83.3\,\% of the studied population), and 100 with all socio-demographic information (72.5\,\% of the studied population).

For ICCSS17, there were 339 participants, from which 274 accepted to take part in the present study (80.8\,\%). The conference consisted on a workshop day and 3 days of conference. We restrict ourselves to the conference days, for which we have 262 participants to the study. Among them, we have at least partial socio-demographic information for 202 participants (77.1\,\% of the studied population), and 188 with all socio-demographic information (71.8\,\% of the studied population).

See the Supplementary Information for a complete description of the data set.

Data sets are available on Zenodo at the following address: \url{https://zenodo.org/record/2531537}

\subsection{Contact matrices}

A connectivity contact matrix is a square matrix of the local densities of links between and within each group, as defined by the socio-demographic attributes. The local density $\rho_{ij}$ is defined as the following:
\begin{equation}\label{eq:rho}
  \rho_{ij} = \dfrac{E_{ij}}{N_{ij}}
\end{equation}
where $E_{ij}$ is the number of existing links between groups $i$ and $j$, and $N_{ij}$ is the number of possible links between these two groups. $N_{ij}$ is given by:
\begin{equation}\label{eq:density}
  N_{ij} = \left\{
    \begin{array}{cr}
      n_i n_j & \text{between two groups}\\
      \\
      \dfrac{n_i(n_i - 1)}{2} & \text{within one group}\\
    \end{array}
  \right.
\end{equation}
where $n_i$ is the number of individuals in the group $i$. This measure is thus independent of the sizes of the groups.

A contact matrix in interactivity is a square matrix of the average link weight between and within each group, as defined by the socio-demographic attributes. Each element of the matrix is defined as the following:
\begin{equation}
  \label{eq:av_weight}
  W_{ij} = \dfrac{\sum_{\ell \in \Omega_{ij}} w_\ell}{E_{ij}}
\end{equation}
where $\Omega_{ij}$ is the set of links between groups $i$ and $j$, $w_\ell$ is the weight of the link $\ell$ and $E_{ij}$ is the number of existing links between groups $i$ and $j$.

\subsection{Null models for the contact matrices}

We here describe the method for assessing the significance of connectivity and interactivity values. The method relies on 1) defining a baseline of network properties, 2) randomising the observed data while retaining the baseline properties, 3) compare the empirical result to the distribution of values generated by the randomisation by computing a deviation score.

\subsubsection{Baselines}

The simplest baseline is to assume that in the absence of all effects, one would get a random, Erd\H{o}s-Renyi network. However, in the present cases we have shown that there exists differences between groups in terms of average degree (as shown in section \ref{sec:group_hetero}). We make the assumption that these differences reflect the actual intrinsic variability of individuals. Therefore, we define baselines that preserve at least individual degrees. Weight distribution is also known to be a key element of face-to-face contact networks, with a particular long-tail shape that is found in every case study \cite{Starnini:PRL2013}. This is the second element that we maintain in all three baselines.

These constraints let us build the three following randomisation methods, ordered from most to least destructive (for the names we use the nomenclature defined in \cite{Gauvin:arxiv2018}):
\begin{itemize}
\item \Pk (Null model 1): also known as a Sneppen-Maslov rewiring \cite{Maslov:Science2002} or configuration model. In this null model all links are randomised, while retaining the degree of each node. Link weights are then redistributed among the new links. The distribution of weights is thus also preserved.
\item \Ppw (Null model 2): in this null model we keep the entire structure of the network (and thus the degree of the nodes), and only redistribute the weights on the links. The distribution of weights is also preserved.
\item \Piso (Null model 3): in this null model we simply randomise the identities of the node in the network. The entire structure is thus preserved, along with the correlations between structure and link weights.
\end{itemize}

These randomisation techniques enable us to test different effects. The first method addresses the question of the particularity of the topology, compared to a situation where connections occur randomly but with the same individual behaviour in terms of connectivity. It allows us to detect if some groups are significantly more or less connected to some other groups. The second method addresses the question of the specificity of the interactions in terms of intensity. Are there pairs of groups which interact significantly more or less than others, compared to a situation with the same connectivity but randomised interactivity? The third method addresses the same issue, but with an additional constraint: are the correlations between topology and interactions group-dependant or not?

\subsubsection{Statistical test}

For each method, we generate 100 randomisations of the original network. For each we compute one or both contact matrices. For \Pk, we compute only the contact matrix in connectivity (as we are interested only in the effect of topology), for \Ppw we compute only the contact matrix in interactivity (as the topology of the network is unchanged), and for \Piso we compute both. We thus generate distributions for each box of the contact matrices, for each method. We then compute the deviations of the observed values from the distributions as a z-score:
\begin{equation}
  z_{ij} = \dfrac{v_{ij}}{\sigma_{ij}}
\end{equation}
where $v_{ij}$ is the contact matrix value for groups $i$ and $j$, and $\sigma_{ij}$ is the standard deviation of the distribution of the contact matrix values for randomised networks for the same pair of groups.

Finally, we assess the significance of these deviations by setting a probability limit $p < 0.01$. However, as values of a contact matrix are not independant, the effective limit depends on the size of the contact matrix:
\begin{equation}
  p_{\rm eff} = \dfrac{p}{(n+1)n/2}
\end{equation}
where $n$ is the number of groups.

\bibliographystyle{unsrt}
\bibliography{biblio}

\begin{thebibliography}{10}

\bibitem{Salathe:PNAS2010}
Marcel Salath{\'e}, Maria Kazandjieva, Jung~Woo Lee, Philip Levis, Marcus~W.
  Feldman, and James~H. Jones.
\newblock A high-resolution human contact network for infectious disease
  transmission.
\newblock {\em Proceedings of the National Academy of Sciences}, 2010.

\bibitem{Sthele:PLOS2011}
Juliette Stehlé, Nicolas Voirin, Alain Barrat, Ciro Cattuto, Lorenzo Isella,
  Jean-François Pinton, Marco Quaggiotto, Wouter Van~den Broeck, Corinne
  Régis, Bruno Lina, and Philippe Vanhems.
\newblock High-resolution measurements of face-to-face contact patterns in a
  primary school.
\newblock {\em PLOS ONE}, 6(8):1--13, 08 2011.

\bibitem{Guclu:PLOS2016}
Hasan Guclu, Jonathan Read, Charles~J. Vukotich, Jr, David~D. Galloway,
  Hongjiang Gao, Jeanette~J. Rainey, Amra Uzicanin, Shanta~M. Zimmer, and Derek
  A.~T. Cummings.
\newblock Social contact networks and mixing among students in k-12 schools in
  pittsburgh, pa.
\newblock {\em PLOS ONE}, 11(3):1--19, 03 2016.

\bibitem{Isella:PLOS2011}
Lorenzo Isella, Mariateresa Romano, Alain Barrat, Ciro Cattuto, Vittoria
  Colizza, Wouter Van~den Broeck, Francesco Gesualdo, Elisabetta Pandolfi,
  Lucilla Ravà, Caterina Rizzo, and Alberto~Eugenio Tozzi.
\newblock Close encounters in a pediatric ward: Measuring face-to-face
  proximity and mixing patterns with wearable sensors.
\newblock {\em PLOS ONE}, 6(2):1--10, 02 2011.

\bibitem{Hornbeck:JID2012}
Thomas Hornbeck, David Naylor, Alberto~M. Segre, Geb Thomas, Ted Herman, and
  Philip~M. Polgreen.
\newblock Using sensor networks to study the effect of peripatetic healthcare
  workers on the spread of hospital-associated infections.
\newblock {\em The Journal of Infectious Diseases}, 206(10):1549--1557, 2012.

\bibitem{Vanhems:PLOS2013}
Philippe Vanhems, Alain Barrat, Ciro Cattuto, Jean-François Pinton, Nagham
  Khanafer, Corinne Régis, Byeul-a Kim, Brigitte Comte, and Nicolas Voirin.
\newblock Estimating potential infection transmission routes in hospital wards
  using wearable proximity sensors.
\newblock {\em PLOS ONE}, 8(9):1--9, 09 2013.

\bibitem{Hertzberg:SN2017}
Vicki~Stover Hertzberg, Jason Baumgardner, C.~Christina Mehta, Lisa~K. Elon,
  George Cotsonis, and Douglas~W. Lowery-North.
\newblock Contact networks in the emergency department: Effects of time,
  environment, patient characteristics, and staff role.
\newblock {\em Social Networks}, 48:181 -- 191, 2017.

\bibitem{Duval:SciRep2018}
Audrey Duval, Thomas Obadia, Lucie Martinet, Pierre-Yves Bo{\"e}lle, Eric
  Fleury, Didier Guillemot, Lulla Opatowski, and Laura Temime.
\newblock Measuring dynamic social contacts in a rehabilitation hospital:
  effect of wards, patient and staff characteristics.
\newblock {\em Scientific reports}, 8(1):1686, 2018.

\bibitem{champredon:AIMS2018}
David Champredon, Mehdi Najafi, Marek Laskowski, Ayman Chit, and Seyed~M
  Moghadas.
\newblock Individual movements and contact patterns in a canadian long-term
  care facility.
\newblock {\em AIMS public health}, 5(2):111, 2018.

\bibitem{Genois:NS2015}
Mathieu G\'enois, Christian~L. Vestergaard, Julie Fournet, André Panisson,
  Isabelle Bonmarin, and Alain Barrat.
\newblock Data on face-to-face contacts in an office building suggest a
  low-cost vaccination strategy based on community linkers.
\newblock {\em Network Science}, 3(3):326–347, 2015.

\bibitem{Montanari:PerCom2017}
A.~Montanari, S.~Nawaz, C.~Mascolo, and K.~Sailer.
\newblock A study of bluetooth low energy performance for human proximity
  detection in the workplace.
\newblock In {\em 2017 IEEE International Conference on Pervasive Computing and
  Communications (PerCom)}, pages 90--99, March 2017.

\bibitem{Kiti:EPJDS2016}
Moses~C. Kiti, Michele Tizzoni, Timothy~M. Kinyanjui, Dorothy~C. Koech,
  Patrick~K. Munywoki, Milosch Meriac, Luca Cappa, Andr{\'e} Panisson, Alain
  Barrat, Ciro Cattuto, and D.~James Nokes.
\newblock Quantifying social contacts in a household setting of rural kenya
  using wearable proximity sensors.
\newblock {\em EPJ Data Science}, 5(1):21, Jun 2016.

\bibitem{Ozella:PLOS2018}
Laura Ozella, Francesco Gesualdo, Michele Tizzoni, Caterina Rizzo, Elisabetta
  Pandolfi, Ilaria Campagna, Alberto~Eugenio Tozzi, and Ciro Cattuto.
\newblock Close encounters between infants and household members measured
  through wearable proximity sensors.
\newblock {\em PLOS ONE}, 13(6):1--16, 06 2018.

\bibitem{Hui:2005PSN}
Pan Hui, Augustin Chaintreau, James Scott, Richard Gass, Jon Crowcroft, and
  Christophe Diot.
\newblock Pocket switched networks and human mobility in conference
  environments.
\newblock In {\em Proceedings of the 2005 ACM SIGCOMM Workshop on
  Delay-tolerant Networking}, WDTN '05, pages 244--251, New York, NY, USA,
  2005. ACM.

\bibitem{Barrat:ISWC2010}
Alain Barrat, Ciro Cattuto, Martin Szomszor, Wouter Van~den Broeck, and Harith
  Alani.
\newblock Social dynamics in conferences: Analyses of data from the live social
  semantics application.
\newblock In Peter~F. Patel-Schneider, Yue Pan, Pascal Hitzler, Peter Mika, Lei
  Zhang, Jeff~Z. Pan, Ian Horrocks, and Birte Glimm, editors, {\em The Semantic
  Web -- ISWC 2010}, pages 17--33, Berlin, Heidelberg, 2010. Springer Berlin
  Heidelberg.

\bibitem{Atzmueller:MMUSM2012}
Martin Atzmueller, Stephan Doerfel, Andreas Hotho, Folke Mitzlaff, and Gerd
  Stumme.
\newblock Face-to-face contacts at a conference: Dynamics of communities and
  roles.
\newblock In Martin Atzmueller, Alvin Chin, Denis Helic, and Andreas Hotho,
  editors, {\em Modeling and Mining Ubiquitous Social Media}, pages 21--39,
  Berlin, Heidelberg, 2012. Springer Berlin Heidelberg.

\bibitem{Isella:JTB2016}
Lorenzo Isella, Juliette Stehlé, Alain Barrat, Ciro Cattuto, Jean-François
  Pinton, and Wouter~Van den Broeck.
\newblock What's in a crowd? analysis of face-to-face behavioral networks.
\newblock {\em Journal of Theoretical Biology}, 271(1):166 -- 180, 2011.

\bibitem{Macek:2012ACM}
Bjoern-Elmar Macek, Christoph Scholz, Martin Atzmueller, and Gerd Stumme.
\newblock Anatomy of a conference.
\newblock In {\em Proceedings of the 23rd ACM Conference on Hypertext and
  Social Media}, HT '12, pages 245--254, New York, NY, USA, 2012. ACM.

\bibitem{Ren:EPJDS2018}
Yongli Ren, Martin Tomko, Flora~D. Salim, Jeffrey Chan, and Mark Sanderson.
\newblock Understanding the predictability of user demographics from
  cyber-physical-social behaviours in indoor retail spaces.
\newblock {\em EPJ Data Science}, 7(1):1, Jan 2018.

\bibitem{Stopczynski:PLOS2014}
Arkadiusz Stopczynski, Vedran Sekara, Piotr Sapiezynski, Andrea Cuttone,
  Mette~My Madsen, Jakob~Eg Larsen, and Sune Lehmann.
\newblock Measuring large-scale social networks with high resolution.
\newblock {\em PLOS ONE}, 9(4):1--24, 04 2014.

\bibitem{Cattuto:PLOS2010}
Ciro Cattuto, Wouter {Van den Broeck}, Alain Barrat, Vittoria Colizza,
  {Jean-François} Pinton, and Alessandro Vespignani.
\newblock Dynamics of person-to-person interactions from distributed {RFID}
  sensor networks.
\newblock {\em PLOS ONE}, 5(7):e11596, 07 2010.

\bibitem{katz1997research}
J~Sylvan Katz and Ben~R Martin.
\newblock What is research collaboration?
\newblock {\em Research policy}, 26(1):1--18, 1997.

\bibitem{toral2011exploratory}
Sergio~L Toral, Nik Bessis, Mar{\'\i}a del~Roc{\'\i}o Martinez-Torres, Florian
  Franc, Federico Barrero, and Fatos Xhafa.
\newblock An exploratory social network analysis of academic research networks.
\newblock In {\em Intelligent Networking and Collaborative Systems (INCoS),
  2011 Third International Conference on}, pages 21--26. IEEE, 2011.

\bibitem{cummings2008collaborates}
Jonathon~N Cummings and Sara Kiesler.
\newblock Who collaborates successfully?: prior experience reduces
  collaboration barriers in distributed interdisciplinary research.
\newblock In {\em Proceedings of the 2008 ACM conference on Computer supported
  cooperative work}, pages 437--446. ACM, 2008.

\bibitem{moody2004structure}
James Moody.
\newblock The structure of a social science collaboration network: Disciplinary
  cohesion from 1963 to 1999.
\newblock {\em American sociological review}, 69(2):213--238, 2004.

\bibitem{abbott2010chaos}
Andrew Abbott.
\newblock {\em Chaos of disciplines}.
\newblock University of Chicago Press, 2010.

\bibitem{lungeanu2015effects}
Alina Lungeanu and Noshir~S Contractor.
\newblock The effects of diversity and network ties on innovations: The
  emergence of a new scientific field.
\newblock {\em American Behavioral Scientist}, 59(5):548--564, 2015.

\bibitem{darbellay2015rethinking}
Fr{\'e}d{\'e}ric Darbellay.
\newblock Rethinking inter-and transdisciplinarity: Undisciplined knowledge and
  the emergence of a new thought style.
\newblock {\em Futures}, 65:163--174, 2015.

\bibitem{labrie2015strategies}
Nanon Labrie, Rebecca Amati, Anne-Linda Camerini, Marta Zampa, and Claudia
  Zanini.
\newblock ``what's in it for us?”´´ six dyadic networking strategies in
  academia.
\newblock {\em Studies in Communication Sciences}, 15(1):158--160, 2015.

\bibitem{melin2000}
G{\"o}ran Melin.
\newblock Pragmatism and self-organization: Research collaboration on the
  individual level.
\newblock {\em Research policy}, 29(1):31--40, 2000.

\bibitem{bourdieu1972esquisse}
Bourdieu Pierre.
\newblock Esquisse d'une th{\'e}orie de la pratique (pr{\'e}c{\'e}d{\'e} de
  trois {\'e}tudes d'ethnologie kabyle), 1972.

\bibitem{bourdieu1980capital}
Pierre Bourdieu.
\newblock Le capital social.
\newblock {\em Actes de la recherche en sciences sociales}, 31(1):2--3, 1980.

\bibitem{coleman1988}
James~S Coleman.
\newblock Social capital in the creation of human capital.
\newblock {\em American journal of sociology}, 94:S95--S120, 1988.

\bibitem{putnam1995}
Robert~D Putnam.
\newblock Bowling alone.
\newblock {\em Journal of Democracy}, pages 65--78, 1995.

\bibitem{wacquant1990sociology}
Lo{\"\i}c~JD Wacquant.
\newblock Sociology as socioanalysis: Tales of homo academicus.
\newblock In {\em Sociological Forum}, volume~5, pages 677--689. Springer,
  1990.

\bibitem{angervall2018academic}
Petra Angervall, Jan Gustafsson, and Eva Silfver.
\newblock Academic career: on institutions, social capital and gender.
\newblock {\em Higher Education Research \& Development}, pages 1--14, 2018.

\bibitem{jungbauer2013determinants}
Monika Jungbauer-Gans and Christiane Gross.
\newblock Determinants of success in university careers: Findings from the
  german academic labor market/erfolgsfaktoren in der wissenschaft--ergebnisse
  aus einer habilitiertenbefragung an deutschen universit{\"a}ten.
\newblock {\em Zeitschrift f{\"u}r Soziologie}, 42(1):74--92, 2013.

\bibitem{bozeman2004capital}
Barry Bozeman and Elizabeth Corley.
\newblock Scientists’ collaboration strategies: implications for scientific
  and technical human capital.
\newblock {\em Research policy}, 33(4):599--616, 2004.

\bibitem{Newman2001structure}
Mark~EJ Newman.
\newblock The structure of scientific collaboration networks.
\newblock {\em Proceedings of the national academy of sciences},
  98(2):404--409, 2001.

\bibitem{lewis2012disciplinarydifferences}
Jenny~M Lewis, Sandy Ross, and Thomas Holden.
\newblock The how and why of academic collaboration: Disciplinary differences
  and policy implications.
\newblock {\em Higher Education}, 64(5):693--708, 2012.

\bibitem{milojevic2012academicage}
Sta{\v{s}}a Milojevi{\'c}.
\newblock How are academic age, productivity and collaboration related to
  citing behavior of researchers?
\newblock {\em PloS one}, 7(11):e49176, 2012.

\bibitem{vanRijnsoever:collaboration}
Frank~J Van~Rijnsoever and Laurens~K Hessels.
\newblock Factors associated with disciplinary and interdisciplinary research
  collaboration.
\newblock {\em Research policy}, 40(3):463--472, 2011.

\bibitem{leydesdorff}
Loet Leydesdorff, Han~Woo Park, and Caroline Wagner.
\newblock International coauthorship relations in the social sciences citation
  index: Is internationalization leading the network?
\newblock {\em Journal of the Association for Information Science and
  Technology}, 65(10):2111--2126, 2014.

\bibitem{zuckerman1972age}
Harriet Zuckerman and Robert~K Merton.
\newblock Age, aging, and age structure in science.
\newblock {\em Higher Education}, 4(2):1--4, 1972.

\bibitem{mertonsociologyofscience}
Robert~K Merton.
\newblock {\em The sociology of science: Theoretical and empirical
  investigations}.
\newblock University of Chicago Press, 1973.

\bibitem{kuhnstructure}
Thomas Kuhn.
\newblock {\em The structure of scientific revolutions. Chicago: Univ}.
\newblock University of Chicago Press, 1962.

\bibitem{friedkin2006structural}
Noah~E Friedkin.
\newblock {\em A structural theory of social influence}, volume~13.
\newblock Cambridge University Press, 2006.

\bibitem{merton1968matthew}
Robert~K Merton.
\newblock The matthew effect in science: The reward and communication systems
  of science are considered.
\newblock {\em Science}, 159(3810):56--63, 1968.

\bibitem{Starnini:PRL2013}
Michele Starnini, Andrea Baronchelli, and Romualdo Pastor-Satorras.
\newblock Modeling human dynamics of face-to-face interaction networks.
\newblock {\em Phys. Rev. Lett.}, 110:168701, Apr 2013.

\bibitem{Gauvin:arxiv2018}
L.~{Gauvin}, M.~{G{\'e}nois}, M.~{Karsai}, M.~{Kivel{\"a}}, T.~{Takaguchi},
  E.~{Valdano}, and C.~L. {Vestergaard}.
\newblock {Randomized reference models for temporal networks}.
\newblock {\em ArXiv e-prints}, June 2018.

\bibitem{Maslov:Science2002}
Sergei Maslov and Kim Sneppen.
\newblock Specificity and stability in topology of protein networks.
\newblock {\em Science}, 296(5569):910--913, 2002.

\end{thebibliography}

\end{document}